\documentstyle[12pt, epsf]{article}

\def \g{\gamma}
\def \d{\delta}

\def \n{\nu}

\def \r{\rho}

\def \s{\sigma}

\def \u{\upsilon}
\def \ph{\phi}

\def \o{\omega}

\def \D{\Delta}

\def \L{\Lambda}

\def \la#1{\label{#1}}

\def \da{\dagger}
\def \ti#1{\tilde{#1}}

\def \rar{\rightarrow}

\def \ld{\ldots}
\def \cd{\cdots}
\def \nn{\nonumber}

\newcommand \beq{\begin{eqnarray}}
\newcommand \eeq{\end{eqnarray}}
\newcommand \beqs{\begin{eqnarray}}
\newcommand \eeqs{\begin{eqnarray}}
\newcommand \ba{\begin{array}}
\newcommand \ea{\end{array}}

\newtheorem{condition}{Condition}
\newtheorem{lemma}{Lemma}
\newtheorem{corollary}{Corollary}
\newtheorem{theorem}{Theorem}
\newtheorem{algorithm}{Algorithm}

\begin{document}
\begin{flushright}
hep-th/9810233\\
\vspace{1cm}
\end{flushright}

\begin{center}
   {\LARGE\bf Poisson Brackets of Normal-Ordered} \\
   \vspace{.2cm}
   {\LARGE\bf Wilson Loops} \\
   \vspace{2cm}
   {\large\bf C.-W. H. Lee and  S. G. Rajeev} \\
   {\it Department of Physics and Astronomy, P.O. Box 270171, University of Rochester, 
    Rochester, New York 14627} \\
   \vspace{.5cm}
   {October 28, 1998} \\
   \vspace{2cm}
   {\large\bf Abstract}
\end{center}

We formulate Yang--Mills theory in terms of the large-$N$ limit, viewed as a classical limit, of gauge-invariant 
dynamical variables, which are closely related to Wilson loops, via deformation quantization.  We obtain a Poisson 
algebra of these dynamical variables corresponding to normal-ordered quantum (at a finite value of $\hbar$) 
operators.  Comparing with a Poisson algebra one of us introduced in the past for Weyl-ordered quantum operators, 
we find, using ideas closely related to topological graph theory, that these two Poisson algebras are, roughly 
speaking, the same.  More precisely speaking, there exists an invertible Poisson morphism between them.

\begin{flushleft}
\end{flushleft}
\pagebreak

\section{Introduction}
\la{s1}

Among the many different approaches to Yang--Mills theory, one of the most widely studied is the large-$N$ limit
\cite{thwi}.  The large-$N$ limit can be formulated as a classical limit \cite{yaffe}, with a well-defined
phase space and a Poisson bracket between dynamical variables which are functions of the phase space.  The hallmark
of Yang--Mills theory is the gauge invariance of physical observables, and it is natural for us to think that the
dynamical variables should also be gauge-invariant functions.  Next comes naturally this question: is there a 
sensible Poisson bracket between these gauge-invariant functions?  If so, this will be a major step towards the
classical formulation of Yang--Mills theory in the large-$N$ limit.

One of us, together with Turgut, introduced in a previous article \cite{ratu96} such a Poisson bracket.  Consider a 
Yang--Mills theory with matter fields $z^i$, where different matter fields are distinguished by different values of 
the index $i$, in the adjoint representation.  Such a theory can be obtained, for example, by dimensionally reducing
a $D$-dimensional Yang--Mills theory to a 2-dimensional one.  The two color indices carried by the adjoint matter
field can be regarded as matrix entries.  In this sense, the adjoint matter fields are Hermitian matrices.  
Consider the trace of a product of these matrices.  Under a gauge transformation characterized by a unitary matrix
$g$, the adjoint matter fields are changed in the following manner:
\beq 
   z^i \rar g z^i g^{\da}.
\la{2.5.1}
\eeq
Hence the trace remains unchanged, and is thus a gauge-invariant function, and a dynamical variable of the theory.  
We call this gauge-invariant function a loop variable, as this was originally motivated from the study of Wilson 
loops \cite{ratu95}.  

A convenient way to quantize such loop variables is via deformation quantization.  (Deformation quantization was
proposed by Flato, Lichnerowicz and Sternheimer \cite{fllist}.  See also Ref.\cite{hofo}.  Ref.\cite{chpr} gives a 
pedagogical introduction.  A more comprehensive list of references can be found in Ref.\cite{flato}.)  The 
essential idea is that the commutative product of these loop variables is deformed in such a way that when we 
multiply two loop variables, it is as if we are multiplying the two operators they represent. (We say that the loop 
variables are the symbols of these operators.)  As there are different ways to order a product of operators, there 
are also different schemes of deformation quantization.  In Ref.\cite{ratu96}, the operators are Weyl-ordered.  
Then the Poisson bracket of two loop variables can be defined as the large-$N$ limit of the commutator of them.  We 
will review the precise definition of this Poisson bracket at the beginning of Section~\ref{s3}.  In a sense, we 
have obtained a classical limit not by setting $\hbar$ to 0 but by letting $N$ go to infinity.  This Poisson 
bracket dictates the classical dynamics of a system in which the dynamical variables are expressed in terms of 
these loop variables.

However, as most finite-$\hbar$ quantum theory are formulated in terms of normal-ordered operators, it should be
interesting to find another Poisson bracket which corresponds to normal-ordered operators, i.e., the loop variables
should be multiplied in such a way that it is as if we are multiplying normal-ordered operators.  This is the goal
in Section~\ref{s2}.

This Poisson algebra is closely related to the Lie algebras we presented in previous papers \cite{prl, npb, jmp},  
though we derived those Lie algebras in a manner thoroughly independent of this Poisson algebra.  We believe that 
the loop variables have a meaning in noncommutative geometry, and, in some sense, the Lie algebras are linear 
approximations of this Poisson algebra.  We have not yet precisely identified the nature of this approximation, and
this is a subject worthy of being pursued in the future.  Nevertheless, at the end of Section~\ref{s2}, we will 
indicate in a crude manner how the Poisson algebra can be truncated to obtain these Lie algebras.

The next interesting question which comes to mind is: what is the relationship between these two seemingly different
Poisson algebras?  It turns out that when there are only a finite number of distinct Hermitian matrices $z^i$, i.e.,
when $i$ can take on a finite number of distinct values only, these two Poisson brackets are, roughly speaking, the 
same.  More precisely speaking, there exists an invertible Poisson morphism between the two Poisson algebras.  We 
are going to show the existence of this Poisson morphism in Section~\ref{s3}.  The astute reader will notice that 
many of the lemmas in the proof have simple interpretations in terms of topological graph theory.  (For an 
introductory account on topological graph theory, see, e.g., Ref.\cite{grtu}.  However, we will not use any of the 
results there because of the difference between the underlying topologies discussed in that reference and the 
topologies here.)  Indeed, we will derive from first principles some properties of topological graphs which, we 
hope, will be of interest to topological graph theorists.

\section{Deformation quantization}
\la{s2}

We are going to derive a Poisson algebra pertinent to guage theory via deformation quantization in this section.  
Deformation quantization refers to the procedure of defining an algebra of smooth functions in such a way that when
the functions are multiplied, it is as if we are multiplying suitably ordered operators these smooth functions 
represent.  To be more specific, consider the set of all smooth functions on a one-dimensional complex Euclidean 
space.  Let $z$ be a coordinate of this one-dimensional space.  Then we can associate a smooth function 
$f(z, \bar{z})$ on it with a Weyl-ordered operator in the way described by Chari and Pressley \cite{chpr}.  The way 
to associate $f(z, \bar{z})$ with a normal-ordered operator is similar.  Indeed, the first step is to obtain the 
Fourier transform $\hat{f}(\xi,\eta )$ of $f(z,\bar{z})$ first:
\begin{equation}
   \hat{f}(\xi , \eta) = \frac{1}{(2\pi)^{2}} \int dz d\bar{z} 
   f(z,\bar{z}) {\rm e}^{-\frac {\rm i}{\hbar}(\xi z + \eta \bar{z})}.
\label{2.1}
\end{equation}
Here $\xi$ and $\eta$ are still complex variables and $\hbar$ is a quantization parameter. Then the associated 
normal-ordered operator $\Phi (f)$ is defined as:
\begin{equation}
   \Phi (f) = \int d\xi d\eta \hat{f}(\xi , \eta) 
   {\rm e}^{{\rm i} \hbar \xi a^{\dagger}} {\rm e}^{{\rm i} \hbar \eta a}, 
\label{2.2}
\end{equation}
where $a$ and $a^{\dagger}$ are the annihilation and creation operators satisfying 
$\lbrack a, a^{\dagger} \rbrack = 1$ respectively.  We then define a non-commutative associative product 
$\ast_{\hbar}$ such that
\begin{equation}
   \Phi (f_{1} \ast_{\hbar} f_{2} ) = \Phi(f_{1})\Phi(f_{2}).
\label{2.3}
\end{equation}
Eq.(\ref{2.3}) is satisfied if this product is defined as follows:
\begin{equation}
   f_{1} \ast_{\hbar} f_{2} \equiv {\rm e}^{\hbar\frac{\partial}{\partial\bar{z}}
   \frac{\partial}{\partial z'}} f_{1}(z,\bar{z}) f_{2}(z',\bar{z}')|_{z=z', \bar{z}=\bar{z}'}.
\label{2.4}
\end{equation}
Then this operation $\ast_{\hbar}$ is a deformation of the algebra of functions on a one-dimensional complex 
Euclidean space. 

In the physical systems we are interested, the dynamical variables are represented by loop variables.  
Mathematically these loop variables are traces of $N \times N$ matrices.  Thus we would like to generalize the 
above formulation of deformation quantization from ordinary complex variables to $N \times N$ matrices.  
Furthermore, physically each matrix corresponds to a state with a particular set of quantum numbers other than 
color (e.g., momentum).  There are, of course, more than one possible set of quantum numbers and so we would also 
generalize the quantization scheme from a one-dimensional space to a multi-dimensional space.  For the sake of 
simplicity, this dimension is still finite though in the actual physical context it should be infinite. 

Having said this, let us generalize the formulation of deformation quantization to a system of bosons.  Consider a 
complex Euclidean space of dimension $2 \L N^2$, where $\L$ is an arbitrary positive integer. Let $z^i$, where 
$i = - \L$, $- \L + 1$, \ld, -1, 1, 2, \ldots, or $\L$, be a Hermitian $N \times N$ matrix.  An entry of $z^i$ is 
denoted by $z^{ia}_{\;b}$, $a$ and $b$ being the row and column indices respectively.  Denote $z^{-i}$ by 
$\bar{z}^i$.  A {\em normal-ordered loop variable} is a function of the form
\begin{equation}
   \tilde{\phi}^I(z,\bar{z}) = {\rm Tr} z^{i_1} \cdots z^{i_m}
\label{2.5}
\end{equation}
Here $I$ represents the sequence of non-zero integers $i_1, \ldots, i_m$ between $-\L$ and $\L$ inclusive.  
$\tilde{\phi}^I(z, \bar{z})$ is gauge-invariant since it remains unchanged under the gauge transformation given by
Eq.(\ref{2.5.1}).  Linear combinations of products of normal-ordered loop variables form a function space 
${\cal N}$.  Eq.(\ref{2.4}) can be generalized to:
\begin{equation}
   \tilde{\phi}^{I}\ast\tilde{\phi}^{J}(z,\bar{z}) = {\rm e}^{\hbar\gamma^{\mu\nu}\frac{\partial}
   {\partial z^{\mu a}_{\;b}} \frac{\partial}{\partial z'^{\nu b} _{\;a}}} 
   \tilde{\phi}^{I}(z,\bar{z})\tilde{\phi}^{J}(z',\bar{z'})|_{z=z',z^{\dagger}=z'^{\dagger}}
\label{2.6}
\end{equation}
with $\gamma^{\mu\nu}=0$ unless $\mu < 0$ and $\nu > 0$.  In the limit $\hbar\rightarrow\infty$, Eq.(\ref{2.6}) 
produces the ordinary Poisson bracket.

Let us derive from Eq.(\ref{2.6}) a Poisson bracket for a finite value of $\hbar$.  This is done by expanding 
Eq.(\ref{2.6}) as a power series of $\hbar$.  Indeed, we obtain
\begin{eqnarray}
   \tilde{\phi}^I \ast_{\hbar} \tilde{\phi}^J & = & 
   \tilde{\phi}^{I}\tilde{\phi}^{J}+\nonumber \\
   & & \sum_{r=1}^{\infty} \frac{\hbar^{r}}{r!} \gamma^{i_{\mu_{1}}j_{\nu_{1}}}
   \cdots \gamma^{i_{\mu_{r}}j_{\nu_{r}}} \frac{\partial^{r}\tilde{\phi}^{I}}
   {\partial z^{i_{\mu_{1}}a_{1}}_{\;\:b_{1}}\cdots 
   z^{i_{\mu_{r}}a_{r}}_{\;\:b_{r}}} \frac{\partial^{r}\tilde{\phi}^{J}}
   {\partial z^{j_{\nu_{1}}b_{1}}_{\;\:a_{1}}
   \cdots z^{j_{\nu_{r}}b_{r}}_{\;\:a_{r}}},
\label{2.7}
\end{eqnarray}
where $i_{\mu_1}, i_{\mu_2}, \ldots, i_{\mu_r} <0$ and $j_{\nu_1}, j_{\nu_2}, \ldots, j_{\nu_r} > 0$.  We can 
always bring the first set of indices to the order $\mu_1 < \mu_2 < \ldots < \mu_r$ by relabelling the indices. 
Then the set of indices $\nu_1, \nu_2, \ldots , \nu_r$ will be rearranged in one of all $r!$ possible permutations. 
We note that
\begin{equation}
   \frac{\partial\tilde{\phi}^{I}}{\partial z^{ka}_{\;b}}=0
\label{2.8}
\end{equation}
unless $k$ is equal to one of the elements of the loop $(i_{1},\ldots ,i_{m})$.
If $k=i_{\mu}$ for some $\mu =1,\ldots ,m$, then
\begin{equation}
   \frac{\partial\tilde{\phi}^{I}}{\partial z^{ka}_{\;b}} =
   [z^{i_{\mu +1}} z^{i_{\mu +2}} \ldots z^{i_{m}} z^{i_{1}} \ldots z^{i_{\mu -1}}]^b_a.
\label{2.9}
\end{equation}
More generally, when $\mu_{1}<\mu_{2}<\ldots<\mu_{r}$,
\begin{equation}
   \frac{\partial^{r}\tilde{\phi}^{I}}
   {\partial z^{i_{\mu_{1}} a_{1}}_{\;\:b_{1}}
   \partial z^{i_{\mu_{2}} a_{2}}_{\;\:b_{2}} \ldots 
   \partial z^{i_{\mu_{r}} a_{r}}_{\;\:b_{r}}} =
   P^{b_{1}}_{a_{2}}(I(\mu_{1} ,\mu_{2})) P^{b_{2}}_{a_{3}}(I(\mu_{2},\mu{3}))
   \ldots P^{b_{r}}_{a_{1}}(I(\mu_{r},\mu_{1}))
\label{2.10}
\end{equation}
where
\begin{equation}
   P^{b_{1}}_{a_{2}}(I(\mu_{1},\mu_{2}))=\left\{
     \begin{array}{ll}
        [z^{i_{\mu_{1}+1}} z^{i_{\mu_{1}+2}} \ldots z^{i_{\mu_{2}-1}}]^{b_{1}}
        _{a_{2}} & \mbox{if $\mu_{2}>\mu_{1}$} \\
   \lbrack z^{i_{\mu_{1}+1}} z^{i_{\mu_{1}+2}} \ldots z^{i_{m}} z^{i_{1}}\ldots
   z^{i_{\mu_{2}-1}}\rbrack ^{b_{1}}_{a_{2}} & \mbox{if $\mu_{2}<\mu_{1}$}
     \end{array}
     \right.
\label{2.11}
\end{equation}
and so on for the other $P$'s.  Hence, we can substitute Eq.(\ref{2.10}) into Eq.(\ref{2.7}) to get
\begin{eqnarray}
   & & \sum_{\{\sigma\} } \frac{\hbar^{r}}{r!} \gamma^{i_{\mu_{1}} 
   j_{\nu_{\sigma (1)}}} \ldots \gamma^{i_{\mu_{r}} j_{\nu_{\sigma (r)}}} \nonumber \\
   & & \cdot P^{b_{1}}_{a_{2}} (I(\mu_{1},\mu_{2})) P^{b_{2}}_{a_{3}} (I(\mu_{2},\mu_{3})) \ldots \nonumber \\
   & & \cdot P^{b_{r}}_{a_{1}} (I(\mu_{r},\mu_{1})) 
   P^{a_{\sigma (1)}}_{b_{\sigma (2)}} (J(\nu_{\sigma (1)}, \nu_{\sigma (2)}))
   P^{a_{\sigma (2)}}_{b_{\sigma (3)}} (J(\nu_{\sigma (2)}, \nu_{\sigma (3)})) \ldots \nonumber \\
   & & \cdot P^{a_{\sigma (r)}}_{b_{\sigma (1)}} (J(\nu_{\sigma (r)}, \nu_{\sigma (1)}))
\label{2.12}
\end{eqnarray}
for the $r$-th order term.  Here $\sigma$ is any possible permutation of $\nu_{1},\ldots ,\nu_{r}$.

In the large-$N$ limit, the term with the largest number of traces will dominate.  This occurs if the $\nu$ indices 
are in decreasing order up to a cyclic permutation, e.g., $\nu_{2}>\nu_{3}>\ldots >\nu_{r}>\nu_{1}$, etc..  
Then to the first two orders in the large-$N$ limit,
\begin{eqnarray}
   \tilde{\phi}^{I} \ast_{\hbar} \tilde{\phi}^{J} & \simeq & 
   \tilde{\phi}^{I} \tilde{\phi}^{J} + \nonumber \\
   & & \sum_{r=1}^{\infty} \sum_{
   \begin{array}{l}
     \mu_1<\mu_2<\cdots <\mu_r  \\
     (\nu_1>\nu_2>\cdots >\nu_r)
   \end{array}}
   \hbar^{r} \gamma^{i_{\mu_{1}} j_{\nu_{1}}} \cdots
   \gamma^{i_{\mu_{r}} j_{\nu_{r}}} \nonumber \\
   & & \cdot \tilde{\phi}^{I(\mu_{1},\mu_{2})J(\nu_{2},\nu_{1})}
   \tilde{\phi}^{I(\mu_{2},\mu_{3})J(\nu_{3},\nu_{2})} \cdots
   \tilde{\phi}^{I(\mu_{r},\mu_{1})J(\nu_{1},\nu_{r})},
\label{2.13}
\end{eqnarray}
where
\begin{equation}
   \tilde{\phi}^{I(\mu_{1},\mu_{2}) J(\nu_{2},\nu_{1})}=
   P^{b}_{\:a}(I(\mu_{1},\mu_{2})) P^{a}_{\:b}(J(\nu_{2},\nu_{1}))
\label{2.14}
\end{equation}

To ensure that the large-$N$ limit is well defined, we need to normalize the functions $\phi^I$ by some 
$N$-dependent factor.  The normalization is such that the vacuum expectation value of $\phi^I$ remains finite as 
$N \rightarrow \infty$.  Consider the vacuum state of the Hamiltonian $g_{ij} {\rm Tr} z^i \bar{z}^j$, where 
$i, j = 1, \ldots , \L$.  Then the vacuum expectation value of $z^{ia}_{\;b} z^{jc}_{\;d}$ is 
$<z^{ia}_{\;b} z^{jc}_{\;d}>= \gamma^{ij} \delta^a_{\,d} \delta^c_{\,b}$.  Thus the vacuum expectation value of the 
product of an odd number of z's will vanish whereas that of an even number of z's will be given by Wick's theorem.  
A short calculation reveals that the $<\tilde{\phi}^I>$ for the $\phi^I$ defined in Eq.(\ref{2.5}) with $m$ even is 
of order $N^{\:\frac{m}{2}+1}$.  This can further be shown to be independent of the particular form of the 
Hamiltonian.  Consequently, we define the normalized functions:
\begin{equation}
   \phi^I = \frac{1}{N^{\frac{m}{2}+1}} \tilde{\phi}^I.
\label{2.15}
\end{equation}

Combining eqs.(\ref{2.13}) and (\ref{2.14}), we get:
\begin{eqnarray}
   \phi^I \ast_{\hbar} \phi^J & = & \phi^{I}\phi^{J}
   + \frac{1}{N_{c}^{\;2}}\sum_{r=1}^{\infty}\sum_{
   \begin{array}{l}
     \mu_{1}<\mu_{2}<\cdots <\mu_{r} \\
     (\nu_{1}>\nu_{2}>\cdots >\nu_{r})
   \end{array}}
   \hbar^{r}\gamma^{i_{\mu_{1}} j_{\nu_{1}}} \cdots
   \gamma^{i_{\mu_{r}} j_{\nu_{r}}} \nonumber \\
   & & \cdot \phi^{I(\mu_{1},\mu_{2})J(\nu_{2},\nu_{1})}
   \phi^{I(\mu_{2},\mu_{3})J(\nu_{3},\nu_{2})}
   \cdots \phi^{I(\mu_{r},\mu_{1})J(\nu_{1},\nu_{r})}
   \nonumber \\
   & & + O(\frac{1}{N^{\;3}}).
\label{2.16}
\end{eqnarray}
Let us define the Poisson bracket by
\begin{equation}
   \{\phi^I,\phi^J\}_N \equiv \lim_{N\rightarrow\infty} N^2 (\phi^I\ast\phi^J-\phi^J\ast\phi^I).
\label{2.17}
\end{equation}
We then finally obtain
\begin{eqnarray}
   \lefteqn{ \{\phi^{I},\phi^{J}\}_N = } \nn \\
   & & \sum_{r=1}^{\infty}
   \sum_{
   \begin{array}{l}
     \mu_{1}<\mu_{2}<\ldots <\mu_{r} \\
     (\nu_{1}>\nu_{2}>\ldots >\nu_{r})
   \end{array}}
    \hbar^{r} \gamma^{i_{\mu_{1}} j_{\nu_{1}}} \ldots
    \gamma^{i_{\mu_{r}} j_{\nu_{r}}} \nonumber \\
    & & \cdot \phi^{I(\mu_{1},\mu_{2})J(\nu_{2},\nu_{1})}
    \phi^{I(\mu_{2},\mu_{3})J(\nu_{3},\nu_{2})} \ldots
    \phi^{I(\mu_{r},\mu_{1})J(\nu_{1},\nu_{r})} \nonumber \\
    & & - (I\leftrightarrow J)
\label{2.18}
\end{eqnarray}
We can visualize Eq.(\ref{2.18}) by the diagrammatic representations in Fig.~\ref{f1}.

\begin{figure}[ht]
\epsfxsize 5in
\centerline{\epsfbox{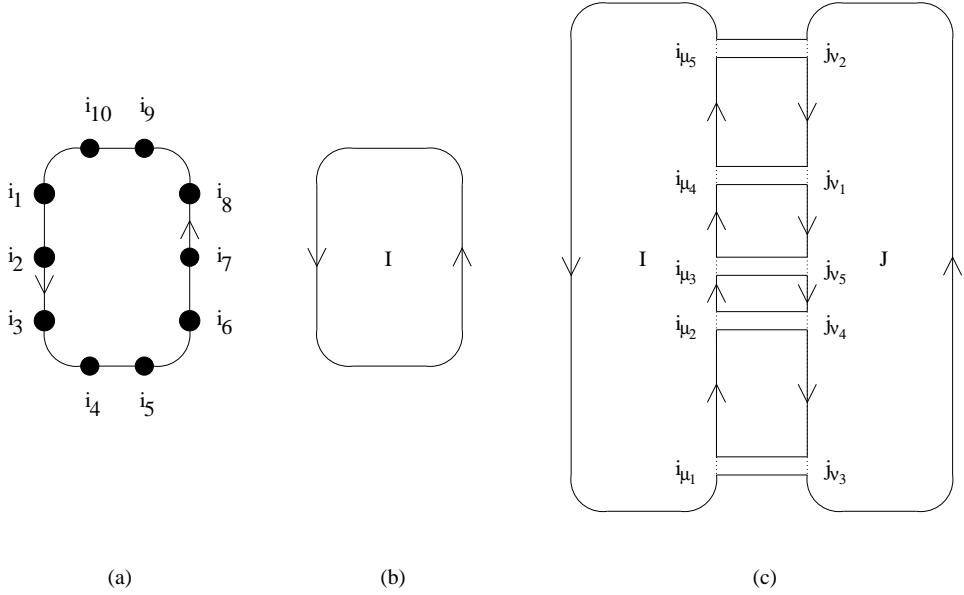}}
\caption{\em (a) A typical loop variable $\ph^I$.  Each solid circle represents a $z^i$.  Notice the cyclic symmetry
of the figure.  (b) A simplified diagrammatic representation of $\ph^I$.  We use the capital letter $I$ to denote 
the whole sequence $i_1$, $i_2$, \ld, $i_m$.  (c) A typical term in $\{ \ph^I, \ph^J \}_N$.  This is a product of 
the loop variables $\ph^{I(\mu_1, \mu_2) J(\n_4, \n_3)}$, $\ph^{I(\mu_2, \mu_3) J(\n_5, \n_4)}$, 
$\ph^{I(\mu_3, \mu_4) J(\n_1, \n_5)}$, $\ph^{I(\mu_4, \mu_5) J(\n_2, \n_1)}$, and 
$\ph^{I(\mu_5, \mu_1) J(\n_3, \n_2)}$.}
\la{f1}
\end{figure}  

Eq.(\ref{2.18}) characterizes the Poisson algebra of loop variables corresponding to normal-ordered operators, and
we call this the {\em normal-ordered Poisson algebra}.  In comparison with the Poisson algebra found in a previous 
paper \cite{ratu96}, where the loop variables correspond to Weyl-ordered operators, we notice that the 
antisymmetric tensors $\omega^{ij}$ in the last equation of Ref.\cite{ratu96} are here replaced by $\gamma^{ij}$, 
which are non-zero only if $i<0$ and $j>0$.  In addition, terms of order $\hbar^1$, $\hbar^3$, $\hbar^5$, \ldots, 
etc. vanish in the last equation of Ref.\cite{ratu96} but they are non-zero here in general.  Nevertheless, these two 
Poisson algebras have a deep relationship --- there is an invertible Poisson morphism between the Poisson algebra 
of Weyl-ordered operators and that of normal-ordered operators, whose proof will be given in the next section.

In previous papers, we defined and discussed a number of Lie algebras like the cyclix algebra \cite{prl}, the 
centrix algebra \cite{npb} and the heterix algebra \cite{jmp}.  These Lie algebras arise from taking the planar
large-$N$ limit of gauge theory.  We can actually think of them as various approximations of the Poisson algebra
given by Eq.(\ref{2.18}).  For example, to get the centrix algebra from this Poisson algebra, we choose $\hbar = 1$
and $\g^{ij} = \d^{-i, j}$ and restrict ourselves to loop variables of the form $\s^I_J \equiv \ph^{IJ^{\ast}} = 
{\rm Tr} z^{i_1} z^{i_2} \cd z^{i_{\#(I)}} \bar{z}^{j_{\#(J)}} \bar{z}^{j_{\#(J)-1}} \cd \bar{z}^{j_1}$, where
$\#(I)$ and $\#(J)$ are the numbers of integers in $I$ and $J$, respectively, and all the indices $i_1$, $i_2$, 
\ld, $i_{\#(I)}$, $j_1$, $j_2$, \ld, and $j_{\#(J)}$ are positive integers between 1 and $\L$ inclusive.  
($J^{\ast}$ is defined as the reverse sequence of $J$.)  If we now compute the Poisson bracket between two loop 
variables of this form using Eq.(\ref{2.18}), we should obtain
\begin{eqnarray}
   \{ \s^I_J, \s^K_L \}_N & = & \sum_{r=1}^{\infty}
   \sum_{
   \begin{array}{l}
     \mu_1 > \mu_2 > \ldots > \mu_r \\
     (\nu_1 > \nu_2 > \ldots > \nu_r)
   \end{array}}
   \phi^{J^{\ast}(\mu_1, \mu_2) K(\nu_2, \nu_1)} \nn \\
   & & \cdot \phi^{J^{\ast}(\mu_2, \mu_3) K(\nu_3, \nu_2)} \cd 
   \phi^{J^{\ast}(\mu_{r-1}, \mu_r) K(\nu_r, \nu_{r-1})} \nn \\
   & & \cdot \phi^{J^{\ast}(\mu_r, 0) I(0, \#(I) + 1) J^{\ast}(\#(J) + 1, \mu_1) K(\nu_r, \#(K) + 1) 
   L^{\ast}(\#(L) + 1, 0) K(0, \nu_1)} \nn \\ 
   & & - (I \leftrightarrow K, J \leftrightarrow L).
\label{2.19}
\end{eqnarray}
If we now retain only those terms in which $\mu_1$, $\mu_2$, \ld, $\mu_r$ are consecutive integers in the reverse 
order, i.e., $\mu_2 = \mu_1 - 1$, $\mu_3 = \mu_2 - 1$, \ld, and $\mu_r = \mu_{r - 1} - 1$, 
and in which $\nu_1$, $\nu_2$, \ld, $\nu_r$ are also consecutive integers in the reverse order, we will get 
precisely the Lie bracket of the centrix algebra.  If we retain some more terms, we will obtain the heterix algebra.
The cyclix algebra is obtained from the heterix algebra by identifying certain products of loop variables as a
linear combination of single loop variables.  We believe that the loop variables have a geometrical meaning in a
noncommutative space, and thus there should be a geometrical meaning of these truncating approximations.  We hope to
understand the geometry better in the future.

\section{A Poisson morphism}
\la{s3}

We are going to show that there exists an invertible Poisson morphism between the Poisson algebra of Weyl-ordered
loop variables described in Ref.\cite{ratu96} and the Poisson algebra of normal-ordered loop variables given in 
Eq.(\ref{2.18}).  It should be interesting for the reader to discern, with the help of the accompanying diagrams,
the meanings of many of the following lemmas in topological graph theory. 

Let us remind ourselves the definition of the Weyl-ordered Poisson algebra here.  Consider $2\L$ distinct 
$N \times N$ Hermitian matrices $\eta^{-\L}$, $\eta^{-\L+1}$, \ldots, $\eta^{-1}$, $\eta^1$, $\eta^2$, \ldots, and
$\eta^{\L}$.  A {\em Weyl-ordered loop variable} is a trace of an arbitrary sequence of these matrices 
$f^I = {\rm Tr} \eta^{i_1} \eta^{i_2} \cdots \eta^{i_m}$, where $m$ is a positive integer called the {\em degree of 
$f^I$}, $i_k \in \{ -M, -M+1, \ldots, -1, 1, 2, \ldots, M \} \; \forall \; k=1, 2, \ldots, m$, and $I$ denotes the 
integer sequence $i_1, i_2, \ldots, i_m$.  Linear combinations of products of Weyl loops form a function space 
$\cal W$.  The Poisson bracket between two Weyl-ordered loop variables $f^I \mbox{and} f^J = {\rm Tr} \eta^{j_1} 
\eta^{j_2} \cdots \eta^{j_n}$, where $n$, $j_k$ and $J$ have analogous definitions as $m$, $i_k$, and $I$, is given 
by the following formula:
\begin{eqnarray}
   \{ f^I, f^J \}_W & = & 2i \sum_{r=1, \mbox{odd}}^{\infty}
   \sum_{\begin{array}{l}
            \mu_1 < \mu_2 < \cdots < \mu_r \\
            (\nu_1 > \nu_2 > \cdots > \nu_r) 
         \end{array}}
   (-\frac{i\hbar}{2})^r \tilde{\omega}^{i_{\mu_1} j_{\nu_1}} \cdots 
   \tilde{\omega}^{i_{\mu_r} j_{\nu_r}} \cdot \nonumber \\
   & & f^{I(\mu_1, \mu_2)J(\nu_2, \nu_1)} f^{I(\mu_2, \mu_3)J(\nu_3, \nu_2)}
   \cdots f^{I(\mu_r, \mu_1)J(\nu_1, \nu_r)}.
\label{3.1}
\end{eqnarray}
In this equation, for every value of $r$, we sum over all possible sets of integers $\mu_1, \mu_2, \ldots, \mu_r 
\in \{ 1, 2, \ldots, m \}$ such that $\mu_1 < \mu_2 < \ldots < \mu_r$, and all sets of integers $\nu_1, \nu_2, 
\ldots$, $\nu_r \in \{ 1, 2, \ldots, m \}$ such that $\nu_1, \nu_2, \ldots, \nu_r$ form a decreasing sequence up to 
a cyclic permutation.  $\tilde{\omega}^{ij}$ is an anti-symmetric tensor.  Eq.(\ref{3.1}) defines the 
{\em Weyl-ordered Poisson algebra} for the space $\cal W$.

Now we are going to define a linear transformation $F:{\cal W} \rightarrow
{\cal N}$.  Nevertheless, we need a number of lemmas first in order to show
that $F$ is well defined.  Introduce two matrices $S$ and $J$ as follows:
\begin{equation}
   S = 2 \left (
   \begin{array}{cc}
      I & 0 \\
      0 & -iI
   \end{array} \right ) \mbox{; and}
\label{3.4}
\end{equation}
\begin{equation}
   J = \left (
   \begin{array}{cc}
      0 & I \\
      -I & 0
   \end{array} \right )
\label{3.5}
\end{equation}
where $I$ is the $\L \times \L$ unit matrix.  The index of each row and column
of $S$ and $J$ runs from 1 to $\L$, then from -1 to $-\L$.  From Eq.(\ref{3.4}),
we see that
\begin{equation}
   S^{-1} = \frac{1}{2} \left(
   \begin{array}{cc}
      I & 0 \\
      0 & iI
   \end{array} \right) .
\label{3.6}
\end{equation}
Let $\eta'^i$, where $i \in \{ -\L, -\L+1, \ldots, -1, 1, 2, \ldots, \L \}$, be defined as 
\begin{equation}
   \eta'^i = (S^{-1})^i (z^i + J^{i \bar{i}} z^{\bar{i}}),
\label{3.7}
\end{equation}
where
\begin{equation}
   (S^{-1})^i = (S^{-1})^{ii}
\label{3.8}
\end{equation}
and
\begin{equation}
   \bar{i} = -i.
\label{3.9}
\end{equation}
Moreover, let
\begin{equation}
   C^{ij} = \gamma^{ij} + J^{i \bar{i}} \gamma^{\bar{i} j} +
   J^{j \bar{j}} \gamma^{i \bar{j}} +
   J^{i \bar{i}} J^{j \bar{j}} \gamma^{\bar{i} \bar{j}}
\label{3.10}
\end{equation}
and
\begin{equation}
   T^{ij} = \frac{\hbar}{2}(S^{-1})^i (S^{-1})^j (C^{ij} + C^{ji}).
\label{3.11}
\end{equation}
We will need these formulae in the definition of $F(f^I)$.

Next we want to introduce the concepts of an allowable set of contracted 
indices, a forbidden set of contracted indices and leftover indices.  Choose
an ordered sequence $I_a$ of $2k$ integers, where $k$ is a non-negative
integer with $2k \leq m$, from $i_1, \ldots, i_m$ with distinct subscripts.
Let us call the integers $i_{a(-1)}$, $i_{a(1)}$, $i_{a(-2)}$, $i_{a(2)}$, \ldots, $i_{a(-k)}$ and $i_{a(k)}$, 
respectively, where $a(-1), a(1), a(-2)$, $a(2), \ldots, 
a(-k), a(k) \in \{ 1, 2, \ldots, m \}$ and $a(r) \neq a(s)$ if $r \neq s$ for 
integers $r$ and $s$ such that $1 \leq \mid r \mid \leq k$ and $1 \leq \mid s 
\mid \leq k$.  Then $I_a = (i_{a(-1)}, i_{a(1)}, i_{a(-2)}, i_{a(2)}, i_{a(-k)}, i_{a(k)})$ 
will be called {\em an allowable set of contracted indices} (or in short $I_a$
is {\em allowable}) if any arbitrary integers $r$ and $s$ such that $1 \leq r < s \leq k$, 
\begin{condition}
either $i_{a(\pm s)} \in I(a(-r), a(r))$ or $i_{a(\pm s)} \in I(a(r),$ $a(-r))$.
\end{condition}
Otherwise, $I_a$ will be a {\em forbidden set of contracted indices} (or in
short $I_a$ is {\em forbidden}).  We illustrate in Fig.~\ref{f2} examples of an allowable set of contracted 
indices, and one of a forbidden set of contracted indices.

\begin{figure}[ht]
\epsfxsize 5in
\centerline{\epsfbox{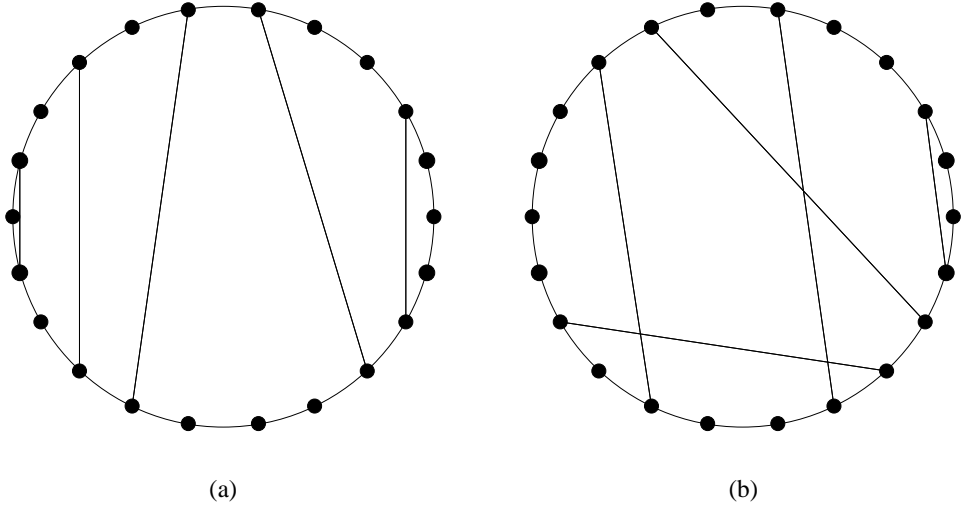}}
\caption{\em (a) An allowable set of contracted indices in a loop variable (Weyl-ordered or normal-ordered).  Each
straight line joins $i_{a(-r)}$ and $i_{a(r)}$ together.  Note that no two straight lines cross each other.  (b)
A forbidden set of contracted indices.  Note that some straight lines cross one other.}
\la{f2}
\end{figure}

We have the following lemmas characterizing an allowable set of contracted indices.

\begin{lemma}
Let $I_a$ be an allowable set of $2k$ contracted indices, and $r$ any integer
between 1 and $k$ inclusive.  Let $b(-1), b(1), b(-2), b(2), \ldots, b(-k),
b(k)$ be an ordered sequence of integers such that for each r, either $b(\pm r)
= a(\pm r)$ or $b(\pm r) = a(\mp r)$.  Then $I_b = (i_{b(-1)}, i_{b(1)},
i_{b(-2)}, i_{b(2)}, \ldots, i_{b(-k)}, i_{b(k)})$ is also an allowable set of 
contracted indices.  If $I_a$ is forbidden, then $I_b$ is also forbidden.
\label{l1}
\end{lemma}
{\bf Proof}. Trivial.

\begin{lemma}
Let $I_a = (i_{a(-1)}, i_{a(1)}, i_{a(-2)}, i_{a(2)}, \ldots, i_{a(-k)},
i_{a(k)})$ be an allowable set of contracted indices, $p$ an arbitrary integer 
between 1 and $k-1$ inclusive, $r$ any integer between 1 and $k$ inclusive, and
$(b(-1)$, $b(1)$, $b(-2)$, $b(2)$, \ldots, $b(-k)$, $b(k))$ a sequence of integers such 
that
\begin{equation}
  \left \{\begin{array}{l}
     b(\pm r) = a(\pm r) \mbox{ if $r \neq p$ and $r \neq p+1$;} \\
     b(\pm p) = a(\pm (p+1)) \mbox{; and} \\
     b(\pm (p+1)) = a(\pm p).
  \end{array} \right.
\label{12}
\end{equation}
Then $I_b = (i_{b(-1)}, i_{b(1)}, i_{b(-2)}, i_{b(2)}, \ldots, i_{b(-k)},
i_{b(k)})$ is also an allowable set of contracted indices.  If $I_a$ is
forbidden, then $I_b$ is also forbidden.
\label{l2}
\end{lemma}
{\bf Proof}. Assume that $I_a$ is allowable.  From Lemma \ref{l1}, we can assume without loss of generality that 
$a(-p) < a(p)$.  It is clear that the set of integers $(i_{b(-1)}, i_{b(1)}, i_{b(-2)}, i_{b(2)}, \ldots, 
i_{b(-p)}, i_{b(p)})$ satisfies Condition~1.  Consider $i_{b(-p-1)}$ and $i_{(p+1)}$.  Since $I_a$ is allowable, we 
have from Condition 1 that either $i_{b(\pm (p+1))} = i_{a(\pm p)} \in I(a(-s), a(s)) = I(b(-s), b(s))$ or 
$i_{b(\pm (p+1))} \in I(b(s), b(-s)) \; \forall \; s = 1, 2, \ldots, p-1$.  Moreover, we have either case~(1) that 
$i_{b(\pm p)} = i_{a(\pm (p+1))} \in I(a(-p), a(p)) = I(b(-p-1), b(p+1))$; case~(2) that 
$i_{b(\pm p)} < i_{b(-p-1)}$; case~(3) that $i_{b(\pm p)} > i_{b(p+1)}$; or case~(4) that $i_{b(-p)} < i_{b(-p-1)}$ 
and $i_{b(p)} > i_{b(p+1)}$.  If one of the first 3 cases holds, then $i_{b(\pm (p+1))} \in I(b(p), b(-p))$.  If 
case~(4) holds, then $i_{b(\pm (p+1))} \in I(b(-p), b(p))$.  Hence in all cases, the set of integers $(i_{b(-1)}, 
i_{b(1)}, i_{b(-2)}, i_{b(2)}, \ldots, i_{b(-p-1)}, i_{b(p+1)})$ satisfies Condition~1.  It is now easy to deduce
that the whole set $(i_{b(-1)}$, $i_{b(1)}$, $i_{b(-2)}$, $i_{b(2)}$, \ld, $i_{b(-k)}$, $i_{b(k)})$ satisfies 
Condition~1.  Hence $I_b$ is also allowable.  The proof that $I_b$ is forbidden if $I_a$ is forbidden is similar. 
Q.E.D.

\begin{lemma}
Consider an allowable set of contracted indices 
\[ I_a = (i_{a(-1)}, i_{a(1)}, i_{a(-2)}, i_{a(2)}, \ldots, i_{a(-k)}, i_{a(k)}). \]  
Let $\sigma : \{ 1, 2, \ldots, k \} \rightarrow \{ 1, 2, \ldots, k \}$ be a permutation of the set of integers 1, 
2, \ldots, and $k$.  Then 
\[ I_{\sigma (a)} = (i_{a(-\sigma (1))}, i_{a(\sigma (1))}, i_{a(-\sigma (2))}, i_{a(\sigma (2))}, \ldots, 
i_{a(-\sigma (k))}, i_{a(\sigma (k))}) \] 
is also allowable.  If $I_a$ is forbidden, then $I_{\sigma (a)}$ is also forbidden.
\label{l3}
\end{lemma}
{\bf Proof}. This can be easily deduced from Lemma~\ref{l2}.
Q.E.D.

\medskip
In short, we see from Lemma~\ref{l3} that whether a set of contracted indices
$I_a$ is allowable or not is independent of the order of the pairs of indices
$i_{a(s)}, i_{a(-s)}$'s.  Each of these pairs will be called a {\em contraction
pair}.

Let us concentrate on an allowable set of contracted indices $I_a$.  For the
$i_l$'s such that $l \in \{ 1, 2, \ldots, m \}$ but that $l \neq a(s) \in
I_a \; \forall \; s = \pm 1, \ldots, \pm k$ (these $i_l$'s are called the {\em
leftover indices}), form {\em subloops} by defining an integer-valued 
auxiliary function $L$ of some positive integers as below.  Let $L(1) = l$. If
$L(\u)$ is defined for an integer $\upsilon$, then we define $L^{(i)} (\upsilon
+ 1)$ for some integers $i$ by the following
\begin{algorithm}
(c.f. Fig.~\ref{f3} below)  In the following, $L(\upsilon) +_m 1$ means precisely $L(\upsilon) + 1$ if
$L(\upsilon) \neq m$, and it means 1 if $L(\upsilon) = m$.  Similarly,
$L(\upsilon -_m 1)$ means $L(\upsilon) - 1$ if $L(\upsilon) \neq 1$, and 
it means m if $L(\upsilon) = 1$.
\begin{description}
   \item[Step 1] Set $i=1$.
   \item[Step 2] $L^{(i)}(\upsilon +1) = L(\upsilon) +_m 1$.
   \item[Step 3] If $L^{(i)}(\upsilon +1) \neq a(s) \; \forall \; s = \pm 1, \ldots,
                  \pm k$, then end this algorithm.
   \item[Step 4] Let $s_i$ be such that $a(s_i) = L^{(i)}(\upsilon + 1)$.
   \item[Step 5] Increment the value of $i$ by 1.
   \item[Step 6] Set $L^{(i)} (\upsilon + 1) = a(-s_i) +_m 1$.
   \item[Step 7] Go back to Step 3.
\end{description}
\label{a1}
\end{algorithm}
If $L^{(i)} (\upsilon + 1) \neq L(1)$, where $i$ is the maximum integer such
that $L^{(i)} (\upsilon + 1)$ is defined, then define $L(\upsilon + 1) =
L^{(i)} (\upsilon + 1)$; otherwise, $L(\upsilon + 1)$ and thus $L(\u +2)$, $L(\u + 3)$, \ld, etc. are all left 
undefined.

Before proceeding on using the auxiliary function $L$ to define a subloop, 
we need to show that the above algorithm is well defined by
\begin{lemma}
In the notations of Algorithm~\ref{a1}, if $L(\u)$ is well defined, then there is an $i$ such that 
$L^{(i)}(\upsilon + 1) \neq a(s) \; \forall \; s = \pm 1, \ldots, \pm k$.
\label{l4}
\end{lemma}
{\bf Proof}.  Assume on the contrary that such an $i$ does not exist.  Then
we have an infinite sequence $L^{(1)}(\upsilon + 1), L^{(2)}(\upsilon + 1),
\ldots$ and so on.  Since there are a finite number of $a(s)$'s for $s = \pm 1, \ldots,
\pm k$ only, there is an integer $i_2$ such that $L^{(i_2)}(\upsilon + 1) = L^{(i_1)}(\u + 1) \; \exists \; i_1 \in
\{ 1, 2, \ld, i_2 - 1 \}$.  Consider the case $i_1 
\neq 1$.  Let $L^{(i_1 - 1)} (\upsilon + 1) = a(s_1)$ and $L^{(i_2 - 1)} =
a(s_2)$ for some integers $s_1$ and $s_2$.  Then $L^{(i_1)} (\upsilon + 1) =
a(-s_1) +_m 1$ and $L^{(i_2)}(\upsilon + 1) = a(-s_2) +_m 1$.  Hence $a(-s_1)
+_m 1 = a(-s_2) +_m 1$ and thus $a(s_1) = a(s_2)$, i.e., $L^{(i_1 - 1)}
(\upsilon + 1) = L^{(i_2 - 1)} (\upsilon + 1)$, contradicting the assumption
that $L^{(i_2)}(\upsilon + 1)$ is the first integer that repeats one of the
previous numbers in the sequence.  Now consider the case $i_1=1$.  Then 
$L^{(i_1)}(\upsilon + 1) = L(\upsilon) +_m 1$ and $L^{(i_2)} (\upsilon + 1)
= a(-s_2) +_m 1$.  Hence $L(\upsilon) = a(-s_2)$.  However, $L(\upsilon)$ does
not belong to $I_a$ and this equation is impossible.  Consequently, there is
an $i$ such that $L^{(i)} (\upsilon + 1) \neq a(s)$ for all $s = \pm 1, \ldots,
\pm k$.  Q.E.D.

\medskip
Let $u$ be the maximum integer such that $L(u)$ is defined.  Then the {\em
subloop of $f^I$ with respect to $I_a$} including $i_l$ is given by $\phi^L
= {\rm Tr} \eta'^{i_{L(1)}} \eta'^{i_{L(2)}} \cdots \eta'^{i_{L(u)}}$, where 
$\eta'^i$ is defined in Eq.(\ref{3.7}).  Obviously any one of the leftover
indices belongs to at least one of these subloops.  Moreover, no two distinct
subloops $\phi^{L_1}$ and $\phi^{L_2}$ of $f^I$ with respect to $I_a$ share
even one common $\eta'^{i_l}$ for an arbitrary leftover index $i_l$ because of 
the following two lemmas.
\begin{lemma}
Consider a subloop $\phi^L = {\rm Tr} \eta'^{i_{L(1)}} \cdots \eta'^{i_{L(u)}}$ of
$f^I$ with respect to $I_a$.  Let $r$ and $s$ be integers between 1 and $u$
inclusive.  Then $L(r) \neq L(s)$ if $r \neq s$.
\label{l5}
\end{lemma}
{\bf Proof}.  Assume on the contrary that there exist some integers $\tilde{r}$
and $\tilde{s}$ such that $\tilde{r} \neq \tilde{s}$ but $L(\tilde{r}) =
L(\tilde{s})$.  Choose the smallest integer $r$ out of these $\tilde{r}$ and
$\tilde{s}$'s.  Then $r>1$ from the statement immediately after Algorithm~\ref{a1}.  Let $s$ be the smallest
integer distinct from $r$ such that $L(s) = L(r)$.  Then $s>r>1$.  Again from the statement immediately after
Algorithm~\ref{a1}, there exists an unknown integer $x$ such that $L(r) =
L^{(x)}(r)$.  Consider the following reverse of Algorithm~\ref{a1}:
\begin{algorithm}  Here is the procedure of this algorithm.
\begin{description}
   \item[Step 1] Set $i=0$.
   \item[Step 2] Let an integer $y$ be such that it satisfies the equation
                  $L^{(x-i)}(r) = y +_m 1$.
   \item[Step 3] If $y$ does not belong to $I_a$, then $y=L(r-1)$ from Step~2
                  of Algorithm~\ref{a1} (or else $L^{(x-i)} (r) = y +_m 1$ where
       		  $y \in I_a$ because of Step~6 of Algorithm~\ref{a1}, which is
        	  impossible).  Hence $x - i = 1 \Rightarrow x = i + 1$.  End the algorithm.
   \item[Step 4] Since $y \in I_a$, $L^{(x-i)}(r) = a(-s_{i+1}) +_m 1 \; \exists \;
                  \mbox{integer} \; s_{i+1}$.
   \item[Step 5] Increment the value of $i$ by 1.
   \item[Step 6] From Steps 6 and 4 of Algorithm \ref{a1}, $L^{(x-i)}(r)
                  = a(s_i)$.
   \item[Step 7] Go back to Step 2.
\end{description}
\label{a2}
\end{algorithm}
Hence $L(r-1)$ can be uniquely determined just from the value of $L(r)$ by
Algorithm~\ref{a2}.  Moreover, $L(s-1)$ can be uniquely determined just from the
value of $L(s)$ by the same algorithm.  Since $L(r)=L(s)$, we must have
$L(r-1) = L(s-1)$, contradicting the assumption that $r$ is the smallest
number such that $L(r) = L(s)$ for a number $s>r$. Q.E.D.
\begin{corollary}
The degree of a subloop is a finite positive integer.
\label{c1}
\end{corollary}
{\bf Proof}.  Since the degree of a loop is a finite number only, a subloop
of it also has a finite degree by Lemma~\ref{l5}.
\begin{lemma}
For each distinct $l \in \{ 1, 2, \ldots, m \}$ such that $i_l$ is a leftover
index, $\eta'^{i_l}$ is contained in at most one of the distinct subloops produced from all the leftover 
indices.
\label{l6}
\end{lemma}
{\bf Proof}.  Let $\eta'^{i_l} \in \phi^{L} = {\rm Tr} \eta'^{i_{L(1)}} \cdots
\eta'^{i_{L(u)}}$, where $L(1) = l$.  Consider another subloop $\phi^{\tilde{L}}
= {\rm Tr} \eta'^{i_{\tilde{L}(1)}} \cdots \eta'^{i_{\tilde{L}(\upsilon)}} \cdots
\eta'^{i_{\tilde{L}(\tilde{u})}}$, where $\tilde{L}(\upsilon) = l$.  From 
Algorithm~\ref{a1}, it is clear that $\tilde{L}(\upsilon + 1) = L(2)$, 
$\tilde{L}(\upsilon + 2) = L(3)$, \ldots, and $\tilde{L}(\tilde{u}) = L(\tilde{u}
- \upsilon + 1)$.  Then $\tilde{L}^{(i)}(\tilde{u} + 1) = \tilde{L}(1)$ for the
maximum integer $i$ such that $\tilde{L}^{(i)}(\tilde{u} + 1)$ is defined.  On
the other hand, $\tilde{L}^{(i)}(\tilde{u} + 1) = L^{(i)}(\tilde{u} - \upsilon + 2)
= L(\tilde{u} - \upsilon + 2)$.  Hence $\tilde{L}(1) = L(\tilde{u} - \upsilon
+ 2)$.  Then $\tilde{L}(2) = L(\tilde{u} - \upsilon + 3)$, \ldots, and $\tilde{L}
(u - \tilde{u} + \upsilon - 1) = L(u)$.  $L^{(j)}(u + 1) = L(1)$ for 
the maximum integer $j$ such that $L^{(j)}(u + 1)$ is defined.  However, 
$L^{(j)}(u+1) = \tilde{L}^{(j)}(u - \tilde{u} + \upsilon ) = \tilde{L}
(u - \tilde{u} + \upsilon)$.  As a result, $\tilde{L}(u - \tilde{u} + \upsilon)
= L(1) = \tilde{L}(\upsilon)$.  By Lemma~\ref{l5}, $\tilde{L}(u - \tilde{u}
+ \upsilon) = \tilde{L}(\upsilon)$ only if $u=\tilde{u}$.  Now it is clear
that $\phi^L = \phi^{\tilde{L}}$.  Q.E.D.

\medskip
Fig.~\ref{f3} shows a typical subloop.

\begin{figure}[ht]
\epsfxsize 3in
\centerline{\epsfbox{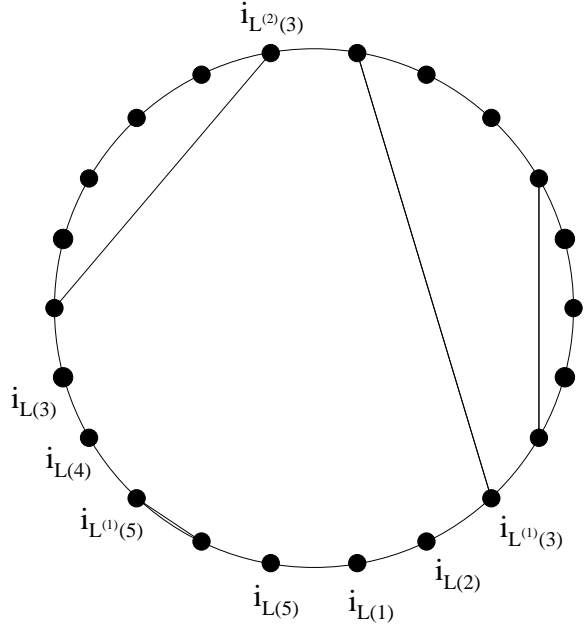}}
\caption{\em A typical subloop with 5 numbers in the integer sequence $L$.}
\la{f3}
\end{figure}

The following lemmas and corollary pertaining to subloops will be found useful later.
\begin{lemma}
Let us consider a particular pair of indices $a(-s_0)$ and $a(s_0)$, where
$1 \leq |s_0| \leq k$,  and the sequence $L^{ext}$ of the numbers
$L(1)$, $L^{(1)}(2)$, $L^{(2)}(2)$, \ldots, $L^{(\iota_2)}(2) = L(2)$, $L^{(1)}(3)$,
$L^{(2)}(3)$, \ldots, $L^{(\iota_{3})}(3) = L(3)$, \ldots, $L^{(1)}(u)$, $L^{(2)}(u)$,
\ldots, $L^{(\iota_u)}(u) = L(u)$, $L^{(1)}(u+1)$, $L^{(2)}(u+1)$, \ldots, and 
$L^{(\iota_{u+1} - 1)}(u+1)$, where $\iota_2$, $\iota_3$, \ldots, $\iota_{u+1}$
are the maximum integers such that $L^{(\iota)}(x)$ is defined for $2 \leq x \leq
u+1$ and $\iota\leq\iota_x$.  Moreover, $L^{(\iota_{u+1})}(u+1) = L(1)$ by the
definition of $u$.  Then either all numbers in $L^{ext} \in I(a(-s_0), a(s_0))
\cup\{ a(s_0) \}$ or all numbers in $L^{ext} \in I(a(s_0), a(-s_0))\cup
\{a(-s_0) \}$.
\label{l7}
\end{lemma}
{\bf Proof}.  Let $L(1) \in I(a(s_0), a(-s_0))\cup\{ a(-s_0) \}$.  Note that
in the sequence $L^{ext}$, the immediately succeeding number $\lambda^{(s)}$
of a preceeding number $\lambda^{(p)}$ is always obtained either by (1) 
$\lambda^{(s)} = \lambda^{(p)} +_m 1$ if $\lambda^{(p)}$ does not belong to
$I_a$, or by (2) $\lambda^{(s)} = a(-s^{(p)}) +_m 1$ for an integer $s^{(p)}$
such that $\lambda^{(p)} = a(s^{(p)})$ if $\lambda^{(p)} \in I_a$.  Assume
that $\lambda^{(p)} \in I(a(s_0), a(-s_0))\cup\{ a(-s_0) \}$.  In Case~(1),
$\lambda^{(p)} \in I(a(s_0), a(-s_0))$ and so $\lambda^{(s)}
\in I(a(s_0), a(-s_0))$ or $\lambda^{(s)} = a(-s_0)$.  Hence $\lambda^{(s)}
\in I(a(s_0), a(-s_0))\cup\{ a(-s_0)\}$.  In Case~(2), if $a(s^{(p)}) \neq
a(-s_0)$, then $a(s^{(p)}) \in I(a(s_0), a(-s_0))$ and so $a(-s^{(p)})\in
I(a(s_0), a(-s_0))$ (because $I_a$ is allowable and because of Lemma~\ref{l3}).
This implies $a(-s^{(p)}) +_m 1 \in I(a(s_0), a(-(s_0))\cup \{ a(-s_0) \}$.
If, on the other hand, $a(s^{(p)}) = a(-s_0)$, then $a(-s^{(p)}) = a(s_0)$ and
so $a(-s^{(p)}) +_m 1 \in I(a(s), a(-s_0)) \cup \{ a(-s_0) \}$.  By induction,
all numbers in $L^{ext} \in I(a(s_0), a(-s_0))\cup\{ a(-s_0)\}$.  The case for
$L(1)\cup I(a(-s_0), a(s_0)) \cup \{ a(s_0) \}$ is similar.  Q.E.D.
\begin{lemma}
Consider a subloop $\phi^{L} = {\rm Tr} \eta'^{i_{L(1)}} \cdots \eta'^{i_{L(u)}}$ of
$f^I$ with respect to $I_a$.  Without loss of generality, $L(1)$ can be chosen
to be the smallest among $L(1), L(2), \ldots, L(u)$ by a cyclic permutation
of the $\eta'$ matrices.  Then using the notations of Lemma~\ref{l5}, we have
$L(1)<L^{(1)}(2)<L^{(2)}(2)<\cdots<L^{(\iota_2)}(2)<L^{(1)}(3)<L^{(2)}(3)
<\cdots<L^{(\iota_3)}(3)<\cdots<L^{(1)}(u)<L^{(2)}(u)<\cdots<L^{(\iota_u)}(u)$.
\label{l8}
\end{lemma}
{\bf Proof}.  Consider the numbers $\lambda^{(p)}$ and $\lambda^{(s)}$ 
defined in the proof of Lemma~\ref{l7}.  Assume that $L(1)<L^{(1)}(2)<\cdots
<L^{(\iota_2)}(2)<\cdots<\lambda^{(p)}$.  If $\lambda^{(p)} = 
L^{(\iota_u)}(u)$, the lemma is proved.  If $\lambda^{(p)}$ is a number before
$L^{(\iota_u)}(u)$ in the sequence $L^{ext}$, then $\lambda^{(s)}$ is obtained
by the two alternatives described in the proof of Lemma~\ref{l7}.  For the case 
$\lambda^{(p)}$ does not belong to $I_a$, we have $\lambda^{(s)} = 
\lambda^{(p)} +_m 1$.  Thus either $\lambda^{(s)} = 1 \leq L(1)$ or
$\lambda^{(s)} > \lambda^{(p)}$.  For the case $\lambda^{(p)} \in I_a$, we
have $\lambda^{(s)} = a(-s^{(p)}) +_m 1$ where $s^{(p)}$ is defined in the proof
of the Lemma~\ref{l7}.  If $a(s^{(p)}) < a(-s^{(p)})$, then either $\lambda^{(s)}
\leq L(1)$ or $\lambda^{(s)} > \lambda^{(p)}$.  If $a(s^{(p)})>a(-s^{(p)})$, then
we deduce from  $\lambda^{(p)} \in I(a(-s^{(p)}), a(s^{(p)})) \cup \{ a(s^{(p)}) \}$ and Lemma~\ref{l7} that
$L(1)\in I(a(-s^{(p)}), a(s^{(p)}))$.  Then 
$\lambda^{(s)} = a(-s^{(p)}) +_m 1 \leq L(1)$.  Hence this lemma is proved if
we can show that $\lambda^{(s)} \leq L(1)$ is impossible.

Clearly, $\lambda^{(s)} = L(1)$ is impossible by Lemma \ref{l5}.  Let
$\lambda^{(s)} < L(1)$, and let $\lambda^{(s)} = L^{(\iota)}(x)$ for some
numbers $\iota$ and $x$.  Consider the numbers $L^{(\iota)}(x)$, 
$L^{(\iota + 1)}(x)$, \ldots, and $L^{(\iota_x)}(x) = L(x)$.  Since $L(x) > L(1)$, 
there is a smallest integer $\iota_c$ such that $L^{(\iota_c)}(x) < L(1)$
but $L^{(\iota_c + 1)}(x) > L(1)$.  Let $L^{(\iota_c)}(x) = a(s_c)$.  Then
$L^{(\iota_c + 1)}(x) = a(-s_c) +_m 1 > L(1)$ and so $a(-s_c) > L(1)$.  This,
together with $a(s_c) < L(1)$, implies $L(1) \in I(a(s_c), a(-s_c))\cup\{
a(-s_c)\}$.  By Lemma~\ref{l7}, $a(s_c) = L^{(\iota_c)}(x) \in 
I(a(s_c), a(-s_c))\cup\{ a(-s_c) \}$, and this is clearly impossible.  Q.E.D.
\begin{corollary}
Consider a subloop $\phi^L = {\rm Tr} \eta'^{i_{L(1)}} \cdots \eta'^{i_{L(u)}}$
of $f^I$ with respect to $I_a$.  $L(1)$ can be chosen to be the smallest
among $L(1), L(2), \ldots, L(u)$ without loss of generality.  Then
$L(1)<L(2)<\cdots<L(u)$.
\label{c2}
\end{corollary}
{\bf Proof}.  This follows directly from Lemmas 6 and 8. Q.E.D.

\medskip
We are now ready to define $F:{\cal W} \rightarrow {\cal N}$.  Let 
$f^I = {\rm Tr} \eta^{i_1}\cdots\eta^{i_m} \in {\cal W}$.  Then
\begin{eqnarray}
   F(f^I) & = & \sum_{\begin{array}{l}
                         \mbox{all distinct allowable sets} \\
                         \mbox{of contracted indices $I_a$}  
                      \end{array}}
   T^{i_{a(-1)} i_{a(1)}} T^{i_{a(-2)} i_{a(2)}} \ldots T^{i_{a(-k)} i_{a(k)}} \nonumber \\
   & & \cdot \prod_{\begin{array}{l}
                 \mbox{all distinct subloops $L$} \\
    		 \mbox{of $f^I$ w.r.t. $I_a$}
              \end{array}}
   {\rm Tr} \eta'^{i_{L(1)}} \eta'^{i_{L(2)}} \cdots \eta'^{i_{L(u)}}.
\label{3.13}
\end{eqnarray}
For instance,
\begin{eqnarray}
   F({\rm Tr} \eta^{i_1} \eta^{i_2} \eta^{i_3} \eta^{i_4}) & = &
   {\rm Tr} \eta'^{i_1} \eta'^{i_2} \eta'^{i_3} \eta'^{i_4} + T^{i_1 i_2} {\rm Tr} \eta'^{i_3} \eta'^{i_4}
   + T^{i_1 i_3} {\rm Tr} \eta'^{i_2} {\rm Tr} \eta'^{i_4} \nn \\
   & & + T^{i_1 i_4} {\rm Tr} \eta'^{i_2} \eta'^{i_3} + T^{i_2 i_3} {\rm Tr} \eta'^{i_1} \eta'^{i_4} +
   T^{i_2 i_4} {\rm Tr} \eta'^{i_1} {\rm Tr} \eta'^{i_3} \nonumber \\
   & & + T^{i_3 i_4} {\rm Tr} \eta'^{i_1} \eta'^{i_2} +
   T^{i_1 i_2} T^{i_3 i_4} + T^{i_1 i_4} T^{i_2 i_3}.
\label{3.14}
\end{eqnarray}
Let us give another example. In $F({\rm Tr} \eta^{i_1} \eta^{i_2} \eta^{i_3} \eta^{i_4} \eta^{i_5} \eta^{i_6})$, 
there are terms like $T^{i_2 i_3} T^{i_5 i_6} {\rm Tr} \eta'^{i_1} \eta'^{i_4}$, 
$T^{i_1 i_6} T^{i_3 i_4} {\rm Tr} \eta'^{i_2} \eta'^{i_5}$ and 
$T^{i_1 i_2} T^{i_4 i_5} {\rm Tr} \eta'^{i_3} \eta'^{i_6}$.  Moreover, we define $F(f^I f^J) = F(f^I) F(f^J)$ for 
some $f^I, f^J \in {\cal W}$.  The following lemma shows that $F$ is invertible.
\begin{lemma}
Consider the mapping $F:{\cal W} \rightarrow {\cal N}$ defined in Eq.(\ref{3.14}).  Then $F$ is invertible.
\label{l9}
\end{lemma}
{\bf Proof}.  Let $P(n)$ be the proposition that for every normal-ordered loop variable $\phi^I$ in ${\cal N}$ of 
degree $n$, there is a unique element in ${\cal W}$ such that $F$ maps this element to $\phi^I$.  From 
Eq.(\ref{3.7}), we see that
\begin{equation}
   \left\{ \begin{array}{l}
      z^j = \eta'^j + {\rm i} \eta'^{-j} \\
      z^{-j} = \eta'^j - {\rm i} \eta'^{-j} 
           \end{array} \right.
\label{3.15}
\end{equation}
for $j \in \{ 1, \ldots, M \}$.  Hence
\begin{equation}
   \left\{ \begin{array}{l}
      {\rm Tr} z^j = {\rm Tr} \eta'^j + {\rm i} \, {\rm Tr} \eta'^{-j} \\
      {\rm Tr} z^{-j} = {\rm Tr} \eta'^j - {\rm i} \, {\rm Tr} \eta'^{-j} 
           \end{array} \right.
\label{3.16}
\end{equation}
for each $j$.  Hence $P(1)$ is true.  Assume that $P(k)$ is true.  Consider 
$\phi^J = {\rm Tr} z^{j_1} \cdots z^{j_{k+1}}$.  From Eq.(\ref{3.15}), $\phi^J$ is a linear combination of 
${\rm Tr} \eta'^{j'_1} \eta'^{j'_2} \cdots \eta'^{j'_{k+1}}$, where $j'_1 = j_1$ or $-j_1$, $j'_2 = j_2$ or $-j_2$,
\ld, and $j'_{k+1} = j_{k+1}$ or $-j_{k+1}$.  Each of these in turn differs from 
$F({\rm Tr} \eta^{j'_1} \eta^{j'_2} \cdots \eta^{j'_{k+1}})$ by normal-ordered loop variables of degrees less than 
$k+1$.  By the induction hypothesis, there is an element $f'$ in {\cal W} which is mapped by $F$ to the sum of 
these normal-ordered loop variables of lower degree.  Hence ${\rm Tr} \eta'^{j'_1} \eta'^{j'_2} \cdots 
\eta'^{j'_{k+1}} = F({\rm Tr} \eta^{j'_1} \eta^{j'_2} \cdots \eta^{j'_{k+1}} + f')$.  Therefore, $P(k+1)$ is true
and there is an element $f^I \in {\cal W}$ such that $F(f^I) = \phi^J$.  If there is another element 
$f^{I'} \neq f^I$ such that $F(f^{I'}) = \phi^J$, then $F(f^{I'} - f^I) = 0$ for $f^{I'} - f^I \neq 0$.  However, 
this is impossbile from Eq.(\ref{3.13}).  Q.E.D.

\medskip
Having defined a mapping $F:{\cal W} \rightarrow {\cal N}$, we are going to prove that
this is a Poisson morphism.  Let $f^I$ and $f^J \in {\cal W}$.  $F$ is a 
Poisson morphism if (1) every term in $\{ F(f^I), F(f^J) \} _N$ is also a term
in $F(\{ f^I, f^J \} _W)$, which is a product of normal-ordered loop variables, and (2) every
term in $F(\{ f^I, f^J \} _W)$ is also a term in $\{ F(f^I), F(f^J) \} _N$.

Let us derive expressions for $\{ F(f^I), F(f^J) \} _N$ and $F(\{ f^I, f^J \} _W)$ first before proving these two 
statements.  From Eq.(\ref{3.13}) and the Leibniz property of a Poisson bracket,
\begin{eqnarray}
   \lefteqn{\{ F(f^I), F(f^J) \} _N = } \nonumber \\
   & & \sum_{\begin{array}{l}
                \mbox{distinct} \\
                \mbox{allowable $I_a$}  
             \end{array}}
   T^{i_{a(-1)} i_{a(1)}} T^{i_{a(-2)} i_{a(2)}} \ldots T^{i_{a(-k)} i_{a(k)}}
   \nonumber \\ 
   & & \cdot \sum_{\begin{array}{l}
		      \mbox{distinct} \\
		      \mbox{allowable $J_b$}  
		   \end{array}}
   T^{j_{b(-1)} j_{b(1)}} T^{j_{b(-2)} j_{b(2)}} \ldots T^{j_{b(-l)} j_{b(l)}} \nonumber \\
   & & \cdot \prod_{\begin{array}{l}
		       \mbox{distinct subloops $L$ of $f^I$} \\
		       \mbox{w.r.t. $I_a$ except subloop $A$}
		    \end{array}}
   {\rm Tr} \eta'^{i_{L(1)}} \eta'^{i_{L(2)}} \cdots \eta'^{i_{L(u)}} \nonumber \\
   & & \cdot \prod_{\begin{array}{l}
		       \mbox{distinct subloops $M$ of $f^J$} \\
		       \mbox{w.r.t. $J_b$ except subloop $B$}
		    \end{array}}
   {\rm Tr} \eta'^{j_{M(1)}} \eta'^{j_{M(2)}} \cdots \eta'^{j_{M(v)}} \nonumber \\
   & & \cdot \{ {\rm Tr} \eta'^{i_{A(1)}} \cdots \eta'^{i_{A(\alpha)}},
   {\rm Tr} \eta'^{j_{B(1)}} \cdots \eta'^{j_{B(\beta)}} \} _N ,
\label{3.17}
\end{eqnarray}
where $J_b = (j_{b(-1)}, j_{b(1)}, j_{b(-2)}, j_{b(2)}, \ldots, j_{b(-l)},
j_{b(l)})$ is an allowable set of contracted indices in $J$ for a postive 
integer $l$, $u$ and $v$ are the degrees of the subloops $L$ and $M$, respectively, and $\alpha$ and $\beta$ are 
the degrees of the subloops $A$ and $B$.  Furthermore, from Eqs.(\ref{3.7}) and (\ref{2.18}),
\begin{eqnarray}
   \lefteqn{\{ {\rm Tr} \eta'^{i_{A(1)}} \cdots \eta'^{i_{A(\alpha)}},
   {\rm Tr} \eta'^{j_{B(1)}} \cdots \eta'^{j_{B(\beta)}} \} _N = } \nonumber \\
   & & \sum_{r'=1}^{\infty} \sum_{\begin{array}{l}
                                    \mu_1 < \mu_2 < \cdots < \mu_{r'} \\
                                    (\nu_1 > \nu_2 > \cdots > \nu_{r'}) 
                                 \end{array}}
   \hbar^{r'} (S^{-1})^{i_{A(\mu_1)}} (S^{-1})^{i_{A(\mu_2)}} \cdots (S^{-1})^{i_{A(\mu_{r'})}} \nonumber \\
   & & \cdot (S^{-1})^{j_{B(\nu_1)}} (S^{-1})^{j_{B(\nu_2)}} \cdots (S^{-1})^{j_{B(\nu_{r'})}} \nonumber \\
   & & \cdot (C^{i_{A(\mu_1)} j_{B(\nu_1)}} C^{i_{A(\mu_2)} j_{B(\nu_2)}} \cdots 
   C^{i_{A(\mu_{r'})} j_{B(\nu_{r'})}} \nonumber \\
   & & - C^{j_{B(\nu_1)} i_{A(\mu_1)}} C^{j_{B(\nu_2)} i_{A(\mu_2)}} \cdots C^{j_{B(\nu_{r'})} i_{A(\mu_{r'})}}) 
   \nonumber \\
   & & \cdot H^{I_A(\mu_1, \mu_2)J_B(\nu_2, \nu_1)} H^{I_A(\mu_2, \mu_3)J_B(\nu_3, \nu_2)}
   \cdots H^{I_A(\mu_{r'}, \mu_1)J_B(\nu_1, \nu_{r'})}. 
\label{3.18}
\end{eqnarray}
In this equation, if $\mu_1 < \mu_2$, then
\begin{equation}
   I_A(\mu_1, \mu_2) = i_{A(\mu_1 + 1)}, i_{A(\mu_1 + 2)}, \ldots, i_{A(\nu_2 - 1)}.
\la{3.19}
\end{equation}
If, instead, $\mu_1 \geq \mu_2$, then
\begin{equation}
   I_A(\mu_1, \mu_2) = i_{A(\mu_1 + 1)}, i_{A(\mu_1 + 2)}, \ldots, i_{A(\alpha)}, i_{A(1)}, i_{A(2)}, \ldots, 
   i_{A(\nu_2 - 1)}.
\label{3.20}
\end{equation}
We have a similar definition for $J_B (\nu_2, \nu_1)$.  In addition,
\begin{equation}
   H^{I_A (\mu_1, \mu_2) J_B (\nu_2, \nu_1)} = {\rm Tr} \eta'^{i_{A(\mu_1 + 1)}}
   \cdots \eta'^{i_{A(\mu_2 - 1)}} \eta'^{j_{B(\nu_2 + 1)}} \cdots 
   \eta'^{j_{B(\nu_1 - 1)}}
\label{3.21}
\end{equation}
if $\mu_1 < \mu_2$ and $\nu_1 > \nu_2$, and so on for $H^{I_A(\mu_2, \mu_3)
J_B(\nu_3, \nu_2)}$, \ldots, etc.

Let us define
\begin{equation}
   \omega^{ij} = {\rm i}(S^{-1})^i (S^{-1})^j (C^{ij} - C^{ji}).
\label{3.23}
\end{equation}
Then Eq.(\ref{3.18}) can be simplified by the following lemma:
\begin{lemma}
(Within the statements and proofs of Lemmas~\ref{l10} and \ref{l11}, 
$i_{A(\mu_k)}$ and $j_{B(\nu_k)}$ will be abbreviated as $i_k$ and $j_k$,
respectively.)  The following identity holds true:
\begin{eqnarray}
   & & (S^{-1})^{i_1} (S^{-1})^{i_2} \cdots (S^{-1})^{i_{r'}} (S^{-1})^{j_1} (S^{-1})^{j_2} \cdots 
   (S^{-1})^{j_{r'}}  \nn \\
   & & \cdot (C^{i_1 j_1} C^{i_2 j_2} \cdots C^{i_{r'} j_{r'}} - C^{j_1 i_1} C^{j_2 i_2} \cdots C^{j_{r'} i_{r'}}) 
   \nonumber \\
   & & = 2 \sum_{\begin{array}{l}
		    \mbox{distinct sets of choices for} \\
		    \mbox{$\Delta$ with an {\em odd} number of $\omega$'s}
		 \end{array}}
   (\Delta)^{i_1 j_1} (\Delta)^{i_2 j_2} \cdots (\Delta)^{i_{r'} j_{r'}}
\label{3.24}
\end{eqnarray}
where each $(\Delta)^{ij}$ can be chosen as either $-\frac{i}{2}\omega^{ij}$ or $\frac{1}{\hbar} T^{ij}$.
\label{l10}
\end{lemma}
In order to prove Lemma \ref{l10}, we need to state and prove Lemma~\ref{l11} simultaneously.
\begin{lemma}
The following identity holds true:
\begin{eqnarray}
   & & (S^{-1})^{i_1} (S^{-1})^{i_2} \cdots (S^{-1})^{i_{r'}} (S^{-1})^{j_1} (S^{-1})^{j_2} \cdots 
   (S^{-1})^{j_{r'}} \nn \\
   & & \cdot (C^{i_1 j_1} C^{i_2 j_2} \cdots C^{i_{r'} j_{r'}} + C^{j_1 i_1} C^{j_2 i_2} \cdots C^{j_{r'} i_{r'}}) 
   \nonumber \\
   & & = 2 \sum_{\begin{array}{l}
		    \mbox{distinct sets of choices of} \\
		    \mbox{$\Delta$ with an {\em even} number of $\omega$'s}
		 \end{array}}
   (\Delta)^{i_1 j_1} (\Delta)^{i_2 j_2} \cdots (\Delta)^{i_{r'} j_{r'}}
\label{3.25}
\end{eqnarray}
\label{l11}
\end{lemma}
{\bf Proof of Lemmas \ref{l10} and \ref{l11}}.  Let us calculate the coefficient of the term
\begin{equation}
   (S^{-1})^{i_1} (S^{-1})^{i_2} \cdots (S^{-1})^{i_{r'}} 
   (S^{-1})^{j_1} (S^{-1})^{j_2} \cdots (S^{-1})^{j_{r'}} 
   C^{i_1 j_1} C^{i_2 j_2} \cdots C^{i_{r'} j_{r'}} 
\label{3.28}
\end{equation}
on the right hand sides of Eqs.(\ref{3.24}) and (\ref{3.25}) first.  This is $\frac{1}{2^{r-1}}$ the number of 
summands on the right hand sides of these two equations because each summand contributes to Formula~\ref{3.28} 
whatever set of choices of $\Delta$'s we choose.  Since the number of distinct choices is
$C^{r'}_1 + C^{r'}_3 + \cdots + C^{r'}_{r'-1}$ or $C^{r'}_1 + C^{r'}_3 + \cdots + C^{r'}_{r'} = 2^{r'-1}$ for 
Eq.(\ref{3.24}) and $C^{r'}_0 + C^{r'}_2 + \cdots + C^{r'}_{r'-1}$ or $C^{r'}_0 + C^{r'}_2 + \cdots + C^{r'}_{r'} = 
2^{r'-1}$ for Eq.(\ref{3.25}), this coefficient is 1.  Similarly, the numerical coefficient of the expression
\begin{equation}
   (S^{-1})^{i_1} (S^{-1})^{i_2} \cdots (S^{-1})^{i_{r'}} 
   (S^{-1})^{j_1} (S^{-1})^{j_2} \cdots (S^{-1})^{j_{r'}} 
   C^{j_1 i_1} C^{j_2 i_2} \cdots C^{j_{r'} i_{r'}} 
\label{3.28.1}
\end{equation}
is -1 on the R.H.S. of Eq.(\ref{3.24}) and 1 on that of Eq.(\ref{3.25}), the negative sign in Eq.(\ref{3.24}) being 
due to the fact that we choose an odd number of the $\Delta$'s to be $\omega$'s, whereas in Eq.(\ref{3.25}) we 
choose an even number.  Hence, every term on the left hand sides of Eq.(\ref{3.24}) and Eq.(\ref{3.25}) are 
contained in the right hand sides of the same equations with the same coefficient.  We are going to show that there 
are no other terms on the R.H.S.'s besides the terms present on the left hand sides.

Indeed, let $C^{<ij>} = C^{ij}$ or $C^{ji}$, and let $P(r', k, -)$ be the proposition that the coefficient of 
$C^{<i_1 j_1>} C^{<i_2 j_2>} \cdots C^{<i_{r'} j_{r'}>}$ where $k$ of the $C^{<ij>}$'s are $C^{ji}$'s and $r'-k$ of 
them are $C^{ij}$'s vanishes on the R.H.S. of Eq.(\ref{3.24}) for $1 \leq k \leq r'-1$.  Similarly, let 
$P(r', k, +)$ be the proposition that this coefficient vanishes on the R.H.S. of Eq.(\ref{3.25}) for $1 \leq k \leq 
r'-1$.  Consider $P(2, 1, -)$ and $P(2, 1, +)$.  The R.H.S. of eq.(\ref{3.24}) is
\begin{eqnarray}
   \lefteqn{ -{\rm i} \frac{T^{i_1 j_1} \omega^{i_2 j_2}}{\hbar} - 
   {\rm i} \frac{\omega^{i_1 j_1} T^{i_2 j_2}}{\hbar} = } \nn \\
   & & (S^{-1})^{i_1} (S^{-1})^{i_2} (S^{-1})^{j_1} 
   (S^{-1})^{j_2} \left( C^{i_1 j_1} C^{i_2 j_2} - C^{j_1 i_1} C^{j_2 i_2} \right).
\label{3.29}
\end{eqnarray}
Therefore, $P(2, 1, -)$ is true.  Similarly, the R.H.S. of eq.(\ref{3.25}) is
\begin{eqnarray}
   \lefteqn{ 2 \frac{T^{i_1 j_1} T^{i_2 j_2}}{\hbar^2} - \frac{1}{2}\omega^{i_1 j_1} \omega^{i_2 j_2} = } \nn \\
   & & (S^{-1})^{i_1} (S^{-1})^{i_2} (S^{-1})^{j_1} (S^{-1})^{j_2} 
   \left( C^{i_1 j_1} C^{i_2 j_2} + C^{j_1 i_1} C^{j_2 i_2} \right).
\label{30}
\end{eqnarray}
Hence $P(2, 1, +)$ is also true.

Now assume that $P(r'', k, -)$ and $P(r'', k, +)$ are true for a positive 
integer $r'' \geq 2$ and $1 \leq k \leq r''-1$.  Consider $P(r'' + 1, k, -)$.  
There are 2 types of summands on the R.H.S. of Eq.(\ref{3.24}) which contributes
to $t=C^{<i_1 j_1>} \cdots C^{<i_{r''} j_{r''}>}$ $C^{<i_{r''+1} j_{r''+1}>}$.
One type ({\em type 1 summands}) is of the general form $(\Delta)^{i_1 j_1}
\cdots$ $(\Delta)^{i_{r''} j_{r''}} T^{i_{r''+1} j_{r''+1}}$. Here an 
odd number of the $\Delta$'s are $\omega$'s, and the rest are $T$'s.  The other
type ({\em type 2 summands}) is of the general form $(\Delta)^{i_1 j_1} \cdots$ 
$(\Delta)^{i_{r''} j_{r''}} \omega^{i_{r''+1} j_{r''+1}}$. Here an even number of
the $\Delta$'s are $\omega$'s.  There are several different cases.

\begin{itemize}
\item Case 1: $2 \leq k \leq r''-1$.
  \begin{itemize}
     \item Subcase a: $t=C^{<i_1 j_1>} \cdots C^{<i_{r''} j_{r''}>} C^{i_{r''+1} j_{r''+1}}$.
	\begin{quotation}
	   Since $P(r'', k, -)$ is true, the coefficient of $t$ derived from type 1 summands, where the first $r''$ 
	   $C^{<ij>}$'s come from $(\Delta)^{ij}$'s and $C^{i_{r''+1} j_{r''+1}}$ comes from 
	   $T^{i_{r''+1} j_{r''+1}}$, is 0.  Since $P(r'', k, +)$ is also true, the coefficient of $t$ derived from 
	   type 2 summands, where $C^{i_{r''+1} j_{r''+1}}$ comes from $\omega^{i_{r''+1} j_{r''+1}}$ instead, is 
	   also 0.  As a result, $P(r''+1, k, -)$ is true in this subcase.
	\end{quotation}
     \item Subcase b: $t=C^{<i_1 j_1>} \cdots C^{<i_{r''} j_{r''}>} C^{j_{r''+1} i_{r''+1}}$.
	\begin{quotation}
	   Since $P(r'', k-1, -)$ is true, the coefficient of $t$ derived from type 1 summands is 0.  Since 
	   $P(r'', k-1, +)$ is also true, the coefficient of $t$ from type 2 summands is also 0.  Hence 
	   $P(r''+1, k, -)$ is also true in this subcase.
	\end{quotation}
  \end{itemize}
\item Case 2: $k=1$.
  \begin{itemize}
     \item Subcase a: $t=C^{<i_1 j_1>} \cdots C^{<i_{r''} j_{r''}>} C^{i_{r''+1} j_{r''+1}}$. 
	\begin{quotation}
	   This is exactly the same as Subcase 1a.
	\end{quotation}
     \item Subcase b: $t=C^{i_1 j_1} \cdots C^{i_{r''} j_{r''}} C^{j_{r''+1} i_{r''+1}}$.
	\begin{quotation}
	   The coefficient of $t$ derived from type 1 summands is $\frac{1}{2}$ (this $\frac{1}{2}$ comes from the 
	   term $\frac{1}{2} C^{j_{r''+1} i_{r''+1}}$ in $T^{i_{r''+1} j_{r''+1}}$), and the coefficient of $t$ 
	   derived from type 2 summands is $-\frac{1}{2}$ (because of the term 
	   $-\frac{1}{2} C^{j_{r''+1} i_{r''+1}}$ in $\omega^{j_{r''+1} i_{r''+1}}$).  Hence the total coefficient 
	   is $\frac{1}{2} - \frac{1}{2} = 0$, i.e., $P(r''+1, 1, - )$ is true in this case.
	\end{quotation}
  \end{itemize}
\item Case 3: $k=r''$
  \begin{itemize}
     \item Subcase a: $t=C^{j_1 i_1} \cdots C^{j_{r''} i_{r''}} C^{i_{r''+1} j_{r''+1}}$.
	\begin{quotation}
	   The coefficient of $t$ derived from type 1 summands is $-\frac{1}{2}$, whereas that derived from type 2 
	   summands is $\frac{1}{2}$.  Hence the total coefficient vanishes and $P(r''+1, r'', -)$ is true in this 
	   case.
	\end{quotation}
     \item Subcase b: $t=C^{<i_1 j_1>} \cdots C^{<i_{r''} j_{r''}>} C^{j_{r''+1} i_{r''+1}}$.
	\begin{quotation}
	   This is the same as Subcase 1b.
	\end{quotation}
  \end{itemize}
\end{itemize}
Thus $P(r''+1, k, -)$ is true for all cases for $1 \leq k \leq r''$.  With
the same induction hypothesis, $P(r''+1, k, +)$ is also true by a similar 
analysis.  By induction, $P(r', k, -)$ and $P(r', k, +)$ are always true
for $r' \geq 2$ and $1 \leq k \leq r'-1$. Q.E.D.

\medskip
With the help of Lemma~\ref{l10}, we can derive from Eqs.(\ref{3.17}) and (\ref{3.18}) that 
$\{ F(f^I), F(f^J) \} _N$ is a linear combination of all terms of the form
\begin{eqnarray}
   & & T^{i_{a(-1)} i_{a(1)}} T^{i_{a(-2)} i_{a(2)}} \cdots T^{i_{a(-k)} i_{a(k)}} T^{j_{b(-1)} j_{b(1)}} 
   T^{j_{b(-2)} j_{b(2)}} \cdots T^{j_{b(-l)} j_{b(l)}} \nonumber \\
   & & \cdot \prod_{\begin{array}{l}
		       \mbox{distinct subloops $L$ of $f^I$} \\
		       \mbox{w.r.t. $I_a$ except subloop $A$}
		    \end{array}}
   {\rm Tr} \eta'^{i_{L(1)}} \eta'^{i_{L(2)}} \cdots \eta'^{i_{L(u)}} \nonumber \\
   & & \cdot \prod_{\begin{array}{l}
		       \mbox{distinct subloops $M$ of $f^J$} \\
		       \mbox{w.r.t. $J_b$ except subloop $B$}
		    \end{array}}
   {\rm Tr} \eta'^{j_{M(1)}} \eta'^{j_{M(2)}} \cdots \eta'^{j_{M(v)}} \nonumber \\
   & & \cdot 2\hbar^{r'} (\Delta)^{i_{A(\mu_1)} j_{B(\nu_1)}} (\Delta)^{i_{A(\mu_2)} j_{B(\nu_2)}} \cdots
   (\Delta)^{i_{A(\mu_{r'})} j_{B(\nu_{r'})}} \nonumber \\
   & & \cdot H^{I_A(\mu_1, \mu_2)J_B(\nu_2, \nu_1)} H^{I_A(\mu_2, \mu_3)J_B(\nu_3, \nu_2)}
   \cdots H^{I_A(\mu_{r'}, \mu_1)J_B(\nu_1, \nu_{r'})} 
\label{3.31}
\end{eqnarray}
with an arbitrary allowable $I_a$, an arbitrary allowable $J_b$, an arbitrary positive integer $r'$, arbitrary sets 
of $\mu$'s and $\nu$'s such that $\mu_1 < \mu_2 < \cdots < \mu_{r'}$ and $(\nu_1 > \nu_2 > \cdots > \nu_{r'})$, and 
an arbitrary set ${\cal C}$ of choices of $\Delta$'s with an odd number of $\omega$'s.  On the other hand, from 
Eqs.(\ref{3.1}) and (\ref{3.13}), $F(\{ f^I, f^J \}) _W)$ is a linear combination of all terms of the form
\begin{eqnarray}
   & & 2 {\rm i} (-\frac{{\rm i}\hbar}{2})^r \omega^{i_{\rho_1} j_{\sigma_1}} \cdots 
   \omega^{i_{\rho_r} j_{\sigma_r}} \nonumber \\
   & & \cdot \prod_{p=1}^r T^{\kappa^{(p)}_{a(-1)} \kappa^{(p)}_{a(1)}} 
   T^{\kappa^{(p)}_{a(-2)} \kappa^{(p)}_{a(2)}} \cdots T^{\kappa^{(p)}_{a(-k)} \kappa^{(p)}_{a(k)}} \nonumber \\
   & & \cdot \prod_{\begin{array}{l}
		       \mbox{distinct subloops $L$} \\ 
		       \mbox{of $f^K$ w.r.t. $K^{(p)}_a$}
		    \end{array}}
   {\rm Tr} \eta'^{\kappa^{(p)}_{L(1)}} \eta'^{\kappa^{(p)}_{L(2)}} \cdots \eta'^{\kappa^{(p)}_{L(u)}}
\label{3.32}
\end{eqnarray}
with an arbitrary positive odd integer $r$, arbitrary sets of $\rho$'s and $\sigma$'s such that $\rho_1 < \rho_2 < 
\cdots < \rho_r$ and $(\sigma_1 > \sigma_2 > \cdots > \sigma_r)$, and an arbitrary allowable set of contracted
indices $K^{(p)}_a$ in
\begin{equation}
   K^{(p)} = \left \{ \begin{array}{l}
       I(\rho_p, \rho_{p+1})J(\sigma_{p+1}, \sigma_p) \; \mbox{for} \;
       1 \leq p \leq r-1 \\
       I(\rho_r, \rho_1)J(\sigma_1, \sigma_r) \; \mbox{for} \; p=r.
   \end{array} \right.
\label{3.33}
\end{equation}
Moreover, the indices of $K^{(p)}$ are $\kappa^{(p)}_1, \kappa^{(p)}_2, \ldots,$ etc..

It is possible to rewrite Eqs.(\ref{3.31}) and (\ref{3.32}) in the same form.  Because of the emergence of a large
number of new parameters, before writing out the new expression (Eq.(\ref{3.34}) below), we would like to introduce 
these parameters first with the help of Fig.~\ref{f4}.

\begin{figure}[ht]
\epsfxsize 5in
\centerline{\epsfbox{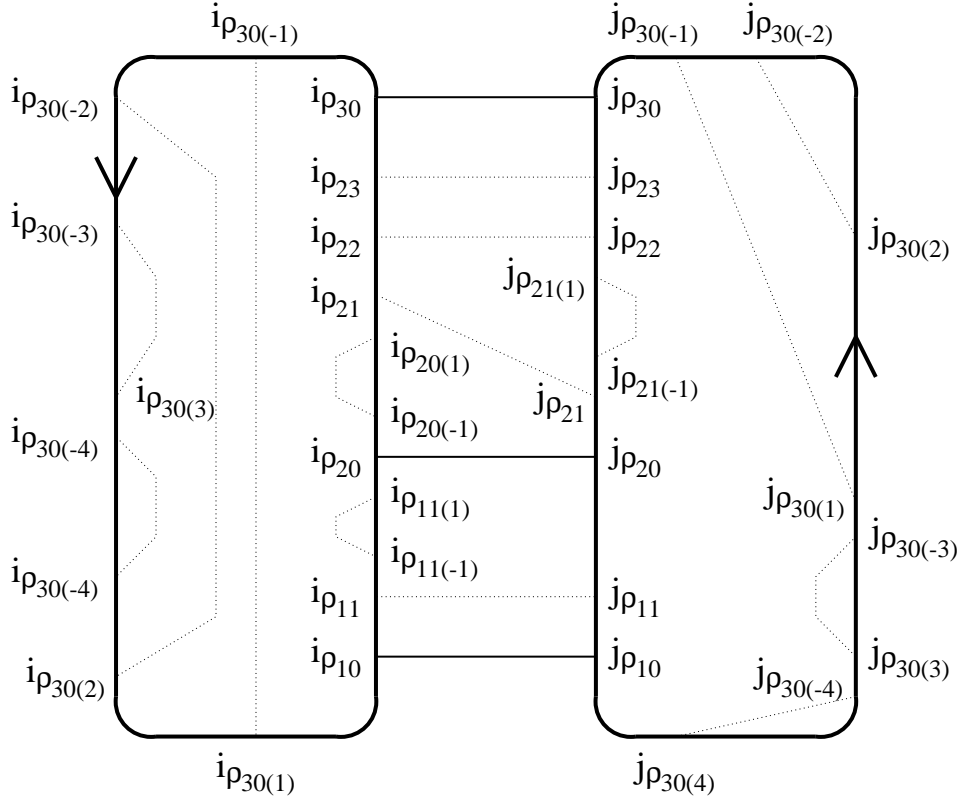}}
\caption{\em An illustration for a typical term in Eq.(\ref{3.34}).  See the text for the legend of this figure.}
\la{f4}
\end{figure}

In Fig.~\ref{f4}, the oval object on the left, which is delineated by a thick closed line with an arrow, is the 
Weyl-ordered loop variable $f^I$.  The oval object on the right is $f^J$.  If we map them to normal-ordered loop
variables and then take the Poisson bracket, we will obtain Eq.(\ref{3.31}); if we take the Poisson bracket first
and map the resultant expression to the space of normal-ordered loop variables later, we will obtain 
Eq.(\ref{3.32}) instead.  There are a number of $\omega^{ij}$'s in both Eqs.(\ref{3.31}) and (\ref{3.32}).  They 
will be labelled as $\o^{i_{\r_{10}} j_{\s_{10}}}$, $\o^{i_{\r_{20}} j_{\s_{20}}}$, \ld, and
$\o^{i_{\r_{r0}} j_{\s_{r0}}}$, where $r$ is a positive {\em odd} integer.  Moreover, it is always possible to
arrange the indices in such a way that $i_{\r_{10}} < i_{\r_{20}} < \cd < i_{\r_{r0}}$ and $(j_{\s_{10}} > 
j_{\s_{20}} > \cd > j_{\s_{r0}})$.  We represent these $\o$'s as solid lines joining the two oval objects in 
Fig.~\ref{f4}.  

There are also a number of $T^{ij}$'s, where the index $i$ comes from $I$ and $j$ comes from $J$, 
in both Eqs.(\ref{3.31}) and (\ref{3.32}).  We will show in the following two lemmas that if they are generically 
labelled as $T^{i_{\r_{x_1 x_2}} j_{\s_{x_1 x_2}}}$, where $1 \leq x_1 \leq r$ and $x_2$ is an integer between 1 
and a certain positive integer $s_{x_1}$, in such a way that
\begin{eqnarray}
   & & (\rho_{10} < \rho_{11} < \cdots < \rho_{1s_1} < \rho_{20} < \rho_{21} < \cdots < \rho_{2s_2} < \cdots \nn \\
   & & < \rho_{r0} < \rho_{r1} < \cdots < \rho_{rs_r}),
\la{3.33.1}
\end{eqnarray}
then
\begin{eqnarray}
   & & (\sigma_{10} > \sigma_{11} > \cdots > \sigma_{1s_1} > \sigma_{20} > \sigma_{21} > \cdots > \sigma_{2s_2} > 
   \cdots \nn \\
   & & > \sigma_{r0} > \sigma_{r1} > \cdots > \sigma_{rs_r}).
\la{3.33.2}
\end{eqnarray}
These $T$'s are depicted as broken lines joining the two oval objects in Fig.~\ref{f4}.  

\la{page1}
There are other $T^{ij}$ in both Eqs.(\ref{3.31}) and (\ref{3.32}) such that both $i$ and $j$ come from $I$.  We 
will also show in the following two lemmas that they can be generically labelled as 
$T^{i_{a_{x_1 x_2 (-x_3)}} i_{a_{x_1 x_2 (x_3)}}}$, where $x_3$ is an integer between 1 and a certain positive 
integer $k_{x_1 x_2}$ in such a way that $i_{a_{x_1 x_2}(\pm x_3)} \in I(\r_{x_1 x_2}, \r_{x_1 x_2 +1})$ for 
$x_2 < s_{x_1}$ or $i_{a_{x_1 x_2}(\pm x_3)} \in I(\r_{x_1 x_2}, \r_{x_1 +_m 1, 0})$ for $x_2 = s_{x_1}$.  These 
$T$'s are depicted as broken lines within the left oval object in Fig.~\ref{f4}.  There are still other $T^{ij}$ in 
both equations such that both $i$ and $j$ come from $J$.  Likewise, we will show that they can be generically 
labelled $T^{j_{b_{x_1 x_2 (-x_3)}} j_{b_{x_1 x_2 (x_3)}}}$, where $x_3$ is an integer between 1 and a certain 
positive integer $l_{x_1 x_2}$ in such a way that $j_{b_{x_1 x_2}(\pm y_3)} \in I(\s_{x_1 x_2}, \s_{x_1 x_2 +1})$ 
for $x_2 < s_{x_1}$ or $j_{b_{x_1 x_2}(\pm y_3)} \in I(\sigma_{x_1 x_2}, \sigma_{x_1 +_m 1, 0})$ for 
$x_2 = s_{x_1}$.  These $T$'s are depicted as broken lines within the right oval object in Fig.~\ref{f4}.

We are now ready to introduce the following lemmas.
\begin{lemma}
   ${\rm i} \{ F(f^I), F(f^J) \} _N$ is equal to a linear combination of all terms of the form
\begin{eqnarray}
   & & 2 {\rm i} \prod_{x_1 = 1}^r (-\frac{{\rm i}\hbar}{2}) \omega^{i_{\rho_{x_1 0}} j_{\sigma_{x_1 0}}} 
   \prod_{x_2 = 1}^{s_{x_1}} T^{i_{\rho_{x_1 x_2}} j_{\sigma_{x_1 x_2}}} \nonumber \\
   & & \cdot \prod_{x_3 = 1}^{k_{x_1 x_2}} T^{i_{a_{x_1 x_2}(-x_3)} i_{a_{x_1 x_2}(x_3)}} 
   \prod_{y_3 = 1}^{l_{x_1 x_2}} T^{\s_{b_{x_1 x_2}(-y_3)} j_{b_{x_1 x_2}(y_3)}} \nonumber \\
   & & \cdot \prod_{\begin{array}{l}
		       \mbox{distinct subloops $L$ within $I$} \\
		       \mbox{w.r.t. all contracted indices in $I$}
		    \end{array}}
   {\rm Tr} \eta'^{i_{L(1)}} \eta'^{i_{L(2)}} \cdots \eta'^{i_{L(u)}} \nonumber \\
   & & \cdot \prod_{\begin{array}{l}
		       \mbox{distinct subloops $M$ within $J$} \\
		       \mbox{w.r.t. all contracted indices in $J$}
		    \end{array}}
   {\rm Tr} \eta'^{j_{M(1)}} \eta'^{j_{M(2)}} \cdots \eta'^{j_{M(v)}} \nonumber \\
   & & \cdot \prod_{\begin{array}{l}
		       \mbox{distinct subloops $QR$ between $I$ and $J$} \\
		       \mbox{w.r.t. all contracted indices in $I$ and $J$}
		    \end{array}}
   {\rm Tr} \eta'^{i_{Q(1)}} \eta'^{i_{Q(2)}} \cdots \eta'^{i_{Q(w_1)}} \nonumber \\ 
   & & \cdot \eta'^{j_{R(1)}} \eta'^{j_{R(2)}} \cdots \eta'^{j_{R(w_2)}},
\label{3.34}
\end{eqnarray}
where 
\begin{enumerate}
   \item $r$ is an arbitrary odd positive integer;
   \item the set of indices $\r_{x_1 x_2}$ is arbitrary except that these indices have to satisfy 
	 Eqs.(\ref{3.33.1}).  Similarly, the set of indices $\s_{x_1 x_2}$ is arbitrary except that these indices
	 have to satisfy Eq.(\ref{3.33.2}).   
   \item the set of all $i_{a_{x_1 x_2}(\pm x_3)}$'s is an arbitrary allowable set of contracted indices in $I$ 
	 satisfying the stipulations in a paragraph on p.\pageref{page1}, and with $i_{a_{x_1 x_2}(-x_3)}$ and 
	 $i_{a_{x_1 x_2}(x_3)}$ forming a contraction pair.  Similarly, the set of all $j_{b_{x_1 x_2}(\pm y_3)}$'s 
	 is an arbitrary allowable set of contracted indices in $J$ also satisfying the stipulations in the same 
	 paragraph, and with $j_{b_{x_1 x_2}(-y_3)}$ and $j_{b_{x_1 x_2}(y_3)}$ forming a contraction pair;
   \item $u$ and $v$ are the degrees of the subloops $L$ and $M$, respectively.  $w_1$ and $w_2$ are integers such 
	 that $w_1 + w_2$ is the degree of the subloop $QR$; and 
   \item the set of all indices in $I$ belonging to any subloop $QR$ comes from one subloop in $I$ with respect to 
	 the set of all $i_{a_{x_1 x_2}(\pm x_3)}$'s.  Similarly, the set of all indices in $J$ belonging to any
	 subloop $QR$ comes from one subloop in $J$ with respect to the set of all $J_{b_{x_1 x_2}(\pm y_3)}$'s.
\end{enumerate}
\label{l12}
\end{lemma}
{\bf Proof}.  First of all, let us prove that any expression of the form shown in Eq.(\ref{3.31}) can be rewritten
in the manner shown in Eq.(\ref{3.34}).  Indeed, let $r$ be the number of $\Delta$'s in Eq.(\ref{3.31}) which are 
$\omega$'s.  $r$ is then a positive odd number (Statement~(1)).  By Corollary~\ref{c2}, $\mu_1 < \mu_2 < \cdots < 
\mu_{r'}$ implies $A(\mu_1) < A(\mu_2) < \cdots < A(\mu_{r'})$ and $(\nu_1 < \nu_2 < \cdots < \nu_{r'})$ implies 
$(B(\nu_1) < B(\nu_2) < \cdots < B(\nu_{r'}))$.  Let us rename the indices 
$A(\mu_1), A(\mu_2), \ldots, A(\mu_{r'})$ and $B(\nu_1) < B(\nu_2), \ldots, B(\nu_{r'})$ by the following
\begin{algorithm} Here is the procedure of this algorithm.
\begin{description}
   \item[Step 1] Set $x_1 = 0$ and $x_2 = 0$.
   \item[Step 2] Set $y=1$.
   \item[Step 3] If $(\Delta)^{i_{A(\mu_y)} j_{B(\nu_y)}} = T^{i_{A(\mu_y)} j_{B(\nu_y)}}$, then increment the 
		 value of $x_2$ by 1.  Put $\rho_{x_1 x_2} = A(\mu_y)$ and $\sigma_{x_1 x_2} = B(\nu_y)$.
   \item[Step 4] If $(\Delta)^{i_{A(\mu_y)} j_{B(\nu_y)}} = \omega ^{i_{A(\mu_y)} j_{B(\nu_y)}}$, then put 
		 $s_{x_1} = x_2$.  Increment the value of $x_1$ by 1 and set the value of $x_2$ to 0.  Put 
		 $\rho_{x_1 0} = A(\mu_y)$ and $\sigma_{x_1 0} = B(\nu_y)$.
   \item[Step 5] If $y \neq r'$, then increment the value of $y$ by 1.  Go back to Step 3.
   \item[Step 6] Put $s_r = x_2 + s_0$.
   \item[Step 7] Set $x'_2 = 1$.
   \item[Step 8] If $x'_2 > s_0$, then end the algorithm.
   \item[Step 9] Put $\r_{r, x_2 + x'_2} = \r_{0, x'_2}$ and $\s_{r, x_2 + x'_2} = \s_{0, x'_2}$.
   \item[Step 10] Increment the value of $x'_2$ by 1.
   \item[Step 11] Go back to Step 8.
\end{description}
\label{a3}
\end{algorithm}
It is now clear that Eqs.(\ref{3.33.1}) and (\ref{3.33.2}) are true (Statement~(2)).  Consider
$i_{a(-s)}$ and $i_{a(s)}$ in Eq.(\ref{3.31}), where $1 \leq \mid s \mid \leq k$.
If $i_{a(-s)} \in I(\rho_{x_1 x_2}, \rho_{x_1, x_2 + 1})$ where $x_2 < s_{x_1}$
but $i_{a(s)} \in I(\rho_{x_1, x_2 + 1}, \rho_{x_1 x_2})$, then $i_{\rho_{
x_1 x_2}} \in I(a(s), a(-s))$ and $i_{\rho_{x_1, x_2 + 1}} \in 
I(a(-s), a(s))$.  Thus two subloops of $f^I$ are involved to produce the 
$\gamma$'s by Lemma~\ref{l7} and then the $\Delta$'s in the Poisson bracket
with $f^J$.  However, only one subloop of $f^I$, namely $A$ in Eq.(\ref{3.31}), 
should be involved and this leads to a contradiction.  Hence, if $i_{a(-s)} \in
I(\rho_{x_1 x_2}, \rho_{x_1, x_2 + 1})$, then $i_{a(s)} \in I(\rho_{x_1 x_2},
\rho_{x_1, x_2 + 1})$ also.  Similarly, if $i_{a(-s)} \in I(\rho_{x_1, x_2 + 1}
, \rho_{x_1 x_2})$, then $i_{a(s)} \in I(\rho_{x_1, x_2 + 1}, \rho_{x_1 x_2})$
also.  In addition, $i_{a(-s)}$ and $i_{a(s)} \in I(\rho_{x_1 s_{x_1}},
\rho_{x_1 +_r 1, 0})$ or $i_{a(-s)}$ and $i_{a(s)} \in I(\rho_{x_1 +_r 1, 0},
\rho_{x_1 s_{x_1}})$.  Let $a_{x_1 x_2}(\pm 1), a_{x_1 x_2}(\pm 2), \ldots$,
$a_{x_1 x_2}(\pm k_{x_1 x_2})$ be those $a(\pm s)$'s such that $a(\pm s) \in
I(\rho_{x_1 x_2}, \rho_{x_1, x_2 + 1})$ for $x_2 < s_{x_1}$ or $a(\pm s) \in
I(\rho_{x_1 s_{x_1}}$, $\rho_{x_1 +_r 1, 0})$ for $x_2 = s_{x_1}$, and let the $b_{x_1 x_2}(\pm x_3)$'s have 
analogous definitions.  These definitions of $a_{x_1 x_2}(x_3)$'s and $b_{x_1 x_2}(y_3)$'s are consistent
with the ones given in the paragraph preceeding this Lemma.  Furthermore, Statement~(3) should be clear.  
Statements~(4) and (5) are direct consequences of Eq.(\ref{3.31}). 

The subloops $L$ and $M$ are still defined by using Algorithm~\ref{a1}.  From
Eqs.(\ref{3.31}) and (\ref{3.34}), every $QR$ lies within $I(\rho_{x_1 x_2},
\rho_{x_1 x_2 + 1})$ and $J(\sigma_{x_1 x_2 + 1}, \sigma_{x_1 x_2})$ for
$x_2 < s_{x_1}$ or within $I(\rho_{x_1 s_{x_1}}, \rho_{x_1 +_r 1, 0})$ and
$J(\sigma_{x_1 s_{x_1}}, \sigma_{x_1 +_r 1, 0})$.  Let $Q(0) = \rho_{x_1
x_2}$.  If $Q(\upsilon)$ is defined for an integer $\upsilon$, then we define
$Q^{(i)}(\upsilon + 1)$ for some integers $i$ as follows:
\begin{algorithm}  Here is the procedure of this algorithm.
\begin{description}
   \item[Step 1] Set $i=1$.
   \item[Step 2] $Q^{(i)}(\upsilon + 1) = Q(\upsilon) +_m 1$.
   \item[Step 3] If $Q^{(i)}(\upsilon + 1) = \rho_{x_1 x_2} \; \exists \; x_1$ and $x_2$, then jump to Step 9.
   \item[Step 4] If $Q^{(i)}(\upsilon + 1) \neq a_{x_1 x_2}(x_3) \; \forall \; x_1, x_2, x_3$ (where $x_3$ can be 
		 positie or negative), then end the algorithm.
   \item[Step 5] Let $\{ x_1, x_2, x_3 \}$ be a set of numbers such that 
		 $a_{x_1 x_2}(x_3) = Q^{(i)}(\upsilon + 1)$.
   \item[Step 6] Increment the value of $i$ by 1.
   \item[Step 7] Set $Q^{(i)}(\upsilon + 1) = a_{x_1 x_2}(-x_3) +_m 1$.
   \item[Step 8] Go back to Step 3.
   \item[Step 9] Let $\{ x_1, x_2 \}$ be a set of numbers such that $\rho_{x_1 x_2} = Q^{(i)}(\upsilon + 1)$.
   \item[Step 10] Increment the value of $i$ by 1.
   \item[Step 11] Set $Q^{(i)}(\upsilon + 1) = \sigma_{x_1 x_2}$.
   \item[Step 12] End the algorithm.
\end{description}
\label{a4}
\end{algorithm}
If the algorithm was ended in Step 4, then define $Q(\upsilon + 1) = Q^{(i)}(\upsilon + 1)$; if the algorithm was 
ended in Step 12, then define $R(0) = Q^{(i)}(\upsilon + 1)$.  If $R(\upsilon)$ is defined for an integer
$\upsilon$, then we define $R^{(i)}(\upsilon + 1)$ for some integers $i$
by {\em Algorithm~5}, which is the same as Algorithm~\ref{a4} except that
$Q^{(i)}(\upsilon + 1)$ is changed to $R^{(i)}(\upsilon + 1)$, $+_m$ to $+_n$,
$\rho_{x_1 x_2}$ to $\sigma_{x_1 x_2}$, $a_{x_1 x_2}(x_3)$ to 
$b_{x_1 x_2}(y_3)$, and $\sigma_{x_1 x_2}$ to $\rho_{x_1 x_2}$.  Then, if Algorithm~5 was ended in Step~4, we will
define $R(\u + 1) = R^{(i)}(\u + 1)$; if it was ended in Step~12, then this $R^{(i)}(\u + 1)$ should be exactly 
$Q(0)$.

Conversely, now let us prove that any expression in the form shown in Eq.(\ref{3.34}) and satisfying the 5 ensuing
statements can be rewritten in the way shown in Eq.(\ref{3.31}).  It should be clear by a reversal of the procedure
described earlier in this proof that we can rewrite the $- {\rm i} \o^{i_{\r_{x_1 0}} j_{\s_{x_1 0}}} / 2$'s
and $T^{i_{\r_{x_1 x_2}} j_{\s_{x_1 x_2}}} / \hbar$'s as $(\D)^{i_{A(\mu_x)} j_{B(\nu_x)}}$'s with $\mu_1 < \mu_2
< \cd < \mu_{r'}$ and $(\n_1 > \n_2 > \cd > \n_{r'})$.  Moreover, the 
$T^{i_{a_{x_1 x_2}(-x_3)} i_{a_{x_1 x_2}(x_3)}}$'s can be rewritten as $T^{i_{a(-x)} i_{a(x)}}$'s, and the
$T^{j_{b_{x_1 x_2}(-y_3)} j_{b_{x_1 x_2}(y_3)}}$'s can be rewritten as $T^{j_{b(-y)} j_{b(y)}}$'s.  If we can show
that the set of all $i_{\r_{x_1 x_2}}$'s come from one subloop with respect to the set of $i_{a_{x_1 x_2}(x_3)}$'s 
in $I$, and the set of all $j_{\s_{x_1 x_2}}$'s come from one subloop with respect to the set of 
$j_{b_{x_1 x_2}(y_3)}$'s in $J$, then the subloops $QR$ can be rewritten as those $H$'s in Eq.(\ref{3.31}).

To show that all $i_{\rho_{x_1 x_2}}$'s come from one subloop, we need to
show that if $L(\upsilon ') = \rho_{x_1 x_2}$ for some values of $x_1$ and
$x_2$, then there exists an integer $\upsilon$ such $L(\upsilon) = \rho_{
x_1 x_2 +1}$ for $x_2 < s_{x_1}$ or $L(\upsilon) = \rho_{x_1 +_r 1, 0}$ for
$x_2 = s_{x_1}$.  We will write $L(\upsilon) = \rho_{x_1 x_2 + 1}$ generically.
We can set $\upsilon ' = 1$ without loss of generality by Lemma~\ref{l6}.
Obviously $(L(1) < \rho_{x_1 x_2 + 1})$.  Assume that $(L(1) < L(2) < \cdots
L(\upsilon - 1) < \rho_{x_1 x_2 + 1})$ for an integer $\upsilon > 1$, and 
assume that $L(\upsilon)$ does not exist.  Then there should be an 
$L^{\iota '}(\upsilon)$ for an integer $\iota ' \geq 1$ such that 
$L^{(\iota ')} (\upsilon) = a_{\tilde{x}_1 \tilde{x}_2}(-\tilde{x}_3)$ for some
integers $\tilde{x}_1, \tilde{x}_2, \tilde{x}_3$ and $a_{\tilde{x}_1
\tilde{x}_2}(\tilde{x}_3) +_m 1 = L(1)$.  Thus $(a_{\tilde{x}_1 \tilde{x}_2}
(\tilde{x}_3) < L(1) < L^{(1)}(\upsilon) < \rho_{x_1 x_2 + 1})$.  Hence
there exists a smallest integer $\iota$ such that $(a_{x_1 x_2}(x_3) < L(1)
< L^{(\iota)}(\upsilon) < \rho_{x_1 x_2 + 1})$, where $a_{x_1 x_2}(-x_3) = 
L^{(\iota)}(\upsilon)$.  However, this is impossible and so $L(\upsilon)$ 
exists.  Assume $(\rho_{x_1 x_2 + 1} < L(\upsilon) < L(1))$.  Then there
should be an $L^{(\iota ')}(\upsilon)$ for an integer $\iota ' \geq 1$ such
that $L^{(\iota ')}(\upsilon) = a_{\tilde{x}_1 \tilde{x}_2}(-\tilde{x}_3)$ 
for some integers $\tilde{x}_1, \tilde{x}_2, \tilde{x}_3$ and
$a_{\tilde{x}_1 \tilde{x}_2}(\tilde{x}_3) +_m 1 = L(\upsilon)$.  Hence
$(L^{(1)}(\upsilon) < \rho_{x_1 x_2 + 1} < a_{\tilde{x}_1 \tilde{x}_2}(x_3)
< L(1))$.  This implies the existence of a smallest integer $\iota$ such
that $L^{(\iota)}(\upsilon) = a_{x_1 x_2}(-x_3)$ for an integer $x_3$ and
$(L^{(\iota)}(\upsilon) < \rho_{x_1 x_2 + 1} < a_{x_1 x_2}(x_3) < L(1))$.
Again this is impossible.  Hence $(L(1) < L(\upsilon) \leq 
\rho_{x_1 x_2 + 1})$.  By Corollary~\ref{c2}, $(L(1) < L(2) < \cdots <
L(\upsilon) \leq \rho_{x_1 x_2 + 1})$.  Since there are only a finite number
of indices between $L(1)$ and $\rho_{x_1 x_2 + 1}$, there exists a number 
$\tilde{\upsilon}$ such that $L(\tilde{\upsilon}) = \rho_{x_1 x_2 + 1}$.
Hence $i_{\rho_{x_1 x_2}}$ and $i_{\rho_{x_1 x_2 + 1}}$ belong to the same
subloop of $I$.  Consequently, the set of all $i_{\rho_{x_1 x_2}}$'s for 
$1 \leq x_1 \leq r$ and $0 \leq x_2 \leq s_{x_1}$  belongs to one subloop
of $I$.  Similarly, the set of all $j_{\sigma_{x_1 x_2}}$'s belongs to one 
subloop of $J$.  Q.E.D.

\medskip
Before we prove Lemma~\ref{l13}, we remark that in the following, by 
$I(\rho_{x_1 0},$ $\rho_{x_1 +_r 1, 0})J(\sigma_{x_1 +_r 1, 0}, \sigma_{x_1 0}) (\rho_{x_1 x_2}, \sigma_{x_1 x_2})$ 
we mean the sequence $i_{\rho_{x_1 x_2} + 1}, i_{\rho_{x_1 x_2} + 2}$, $\ldots, i_{\rho_{x_1 +_r 1, 0}}, 
j_{\sigma_{x_1 +_r 1, 0}}, \ldots, j_{\sigma_{x_1 x_2}}$.
\begin{lemma}
$F(\{ f^I, f^j \} _W)$ can also be written as a linear combination of all terms of the form shown in 
Eq.(\ref{3.34}) with the five accompanying statements.
\label{l13}
\end{lemma}
{\bf Proof}.  First of all, let us show that any expression of the form given in Eq.(\ref{3.32}) can be rewritten
as shown in Eq.(\ref{3.34}) with the accompanying five statements being satisfied.  Indeed, Statement~(1) is 
obvious.  Let us rename the indices $\rho_1$, $\rho_2$, \ldots, $\rho_r$, $\sigma_1$, $\sigma_2$, \ldots, 
$\sigma_r$ in Eq.(\ref{3.32}) as $\rho_{10}$, $\rho_{20}$, \ldots, $\rho_{r0}$, $\sigma_{10}$, $\sigma_{20}$, 
\ldots, $\sigma_{r0}$.  Moreover, within the loop 
$f^{I(\rho_{x_1 0}, \rho_{x_1 +_r 1, 0})J(\sigma_{x_1 +_r 1, 0}, \sigma_{x_1 0})}$, for those contraction pairs 
with one index coming from $I$ and the other one from $J$, call these indices $i_{\rho_{x_1 x_2}}$ and 
$j_{\sigma_{x_1 x_2}}$ in such a way that $\rho_{x_1 0} < \rho_{x_1 1} < \cdots < \rho_{x_1 s_{x_1}} < 
\rho_{x_1 + 1, 0}$ if there are $s_{x_1}$ such pairs for $x_1 < r$, or $(\rho_{r0} < \rho_{r1} < \cdots < 
\rho_{r s_{r}} < \rho_{10})$ if $x_1 = r$.  Thus Eq.(\ref{3.33.1}) is true.  Obviously 
$(\sigma_{x_1 +_r 1, 0} > \sigma_{x_1 1} > \sigma_{x_1 0})$.  Assume that $(\sigma_{x_1 +_r 1, 0} > 
\sigma_{x_1 x_2} > \sigma_{x_1 x_2 -1} > \cdots > \sigma_{x_1 0})$.  If $x_2 \neq s_{x_1}$, consider 
$\sigma_{x_1, x_2 + 1}$.  Since $i_{\rho_{x_1 x_2 + 1}} \in 
I(\rho_{x_1 0}, \rho_{x_1 +_r 1, 0})J(\sigma_{x_1 +_r 1, 0}, \sigma_{x_1 0})(\rho_{x_1 x_2}, \sigma_{x_1 x_2})$, we 
get $j_{\sigma_{x_1, x_2 + 1}} \in I(\rho_{x_1 0}, \rho_{x_1 +_r 1, 0}) 
J(\sigma_{x_1 +_r 1, 0}, \sigma_{x_1 0})(\rho_{x_1 x_2}, \sigma_{x_1 x_2})$ also.  Hence 
$(\sigma_{x_1 +_r 1, 0} > \sigma_{x_1, x_2 + 1} > \sigma_{x_1 x_2} > \sigma_{x_1, x_2 - 1} > \cdots > 
\sigma_{x_1 0})$.  Thus $\sigma_{x_1 +_r 1, 0} > \sigma_{x_1 s_{x_1}} > \sigma_{x_1, s_{x_1} - 1} > \cdots > 
\sigma_{x_1 0}$.  Therefore Statement~(2) is true.  Statements~(3) and (4) are obvious.

Now let us fix the values of $x_1$ and $x_2$, and consider those contraction pairs with both indices coming form 
$I$, and one of them, say $i_c$, belonging to $I(\rho_{x_1 x_2}, \rho_{x_1, x_2 + 1})$ for $x_2 < s_{x_1}$ or
$I(\rho_{x_1 x_2}, \rho_{x_1 +_r 1, 0})$ for $x_2 = s_{x_1}$.  Let $i_{c'}$ be the other index of this contraction 
pair.  Clearly $i_{c'} \in I(\rho_{x_1 0}, \rho_{x_1 +_r 1, 0})$.  If $i_c'$ does not belong to 
$I(\rho_{x_1 x_2}, \rho_{x_1, x_2 + 1})$ for $x_2 < s_{x_1}$ or
$I(\rho_{x_1 x_2}, \rho_{x_1 +_r 1, 0})$ for $x_2 = s_{x_1}$, then 
$i_c \in I(\rho_{x_1 0}, \rho_{x_1 +_r 1, 0})J(\sigma_{x_1 +_r 1, 0}, \sigma_{x_1 0}) 
(\rho_{x_1 x_2}, \sigma_{x_1 x_2})$ but $i_{c'} \in J(\sigma_{x_1 +_r 1, 0}, \sigma_{x_1 0}) 
I(\rho_{x_1 0}, \rho_{x_1 +_r 1, 0})(\sigma_{x_1 x_2}, \rho_{x_1 x_2})$, which is impossible.  Thus any contraction
pairs coming only from $I$ can be written as $i_{a_{x_1 x_2} (\pm x_3)}$ because both of them must belong to a
sequence $I(\r_{x_1 x_2}, \r_{x_1 x_2 + 1})$ with $x_2 < s_{x_1}$, or belong to a sequence 
$I(\r_{x_1 x_2}, \r_{x_1 +_m 1, 0})$ with $x_2 = s_{x_1}$.

We can now say that for a fixed value of $x_1$, the set of all $i_{\rho_{x_1 x_2}}$, $j_{\sigma_{x_1 x_2}}$, 
$i_{a_{x_1 x_2}(\pm x_3)}$ and $j_{b_{x_1 x_2}(\pm y_3)}$, where $1 \leq x_2 \leq s_{x_1}$, $1 \leq x_3 \leq 
k_{x_1 x_2}$ and $1 \leq y_3 \leq l_{x_1 x_2}$, together form an allowable set of contracted indices in the loop
$I(\rho_{x_1 0}, \rho_{x_1 +_r 1, 0}) J(\sigma_{x_1 +_r 1, 0}, \sigma_{x_1 0})$ with $i_{\rho_{x_1 x_2}}$ and 
$j_{\sigma_{x_1 x_2}}$ being contraction pairs, $i_{a_{x_1 x_2}(x_3)}$ and $i_{a_{x_1 x_2}(-x_3)}$ being 
contraction pairs, and $j_{b_{x_1 x_2}(y_3)}$ and $j_{b_{x_1 x_2}(-y_3)}$ being contraction pairs.

The remaining thing to do to prove that Statement~(5) is satisfied is to show that the set of all 
$i_{a_{x_1 x_2}(\pm x_3)}$'s for $1 \leq x_1 \leq r$, $0 \leq x_2 \leq s_{x_1}$ and $1 \leq x_3 \leq k_{x_1 x_2}$ is
allowable in $I$, and the set of all $j_{b_{x_1 x_2}(\pm y_3)}$'s for $1 \leq x_1 \leq r$, 
$0 \leq x_2 \leq s_{x_1}$ and $1 \leq y_3 \leq l_{x_1 x_2}$ is allowable in $J$.  Indeed, since for each fixed 
$x_1$, the set of all $i_{a_{x_1 x_2}(\pm x_3)}$'s is allowable in $I(\rho_{x_1 0}, \rho_{x_1 +_r 1, 0}) 
J(\sigma_{x_1 +_r 1, 0}, \sigma_{x_1 0})$, this set alone is allowable in $I$ also.  Now let us choose a fixed set
of values for $x_1$, $x_2$ and $x_3$.  Consider two integers $\tilde{x}_1$ and $\tilde{x}_2$ such that 
$1 \leq \tilde{x}_1 \leq r$ and $1 \leq \tilde{x}_2 \leq s_{\tilde{x}_1}$.  In addition, either $\ti{x}_1 \neq x_1$ 
or $\ti{x}_2 \neq x_2$, or both.  Then $i_{a_{x_1 x_2}(\pm x_3)} \in I(a_{\tilde{x}_1 \tilde{x}_2}(\tilde{x}_3), 
a_{\tilde{x}_1 \tilde{x}_2}(-\tilde{x}_3))$ for $1 \leq \tilde{x}_3 \leq k_{\tilde{x}_1 \tilde{x}_2}$ if 
$a_{\tilde{x}_1 \tilde{x}_2}(\tilde{x}_3) \geq a_{\tilde{x}_1 \tilde{x}_2}(-\tilde{x}_3)$ and 
$\rho_{\tilde{x}_1 0} < \rho_{\tilde{x}_1 +_r 1, 0}$, or if $a_{\tilde{x}_1 \tilde{x}_2}(\tilde{x}_3) > 
a_{\tilde{x}_1 \tilde{x}_2}(-\tilde{x}_3) > \rho_{\tilde{x}_1 0} > \rho_{\tilde{x}_1 +_r 1, 0}$, or if 
$\rho_{\tilde{x}_1 0} > \rho_{\tilde{x}_1 +_r 1, 0} > a_{\tilde{x}_1 \tilde{x}_2}(\tilde{x}_3) > 
a_{\tilde{x}_1 \tilde{x}_2}(-\tilde{x}_3)$.  Similarly, $i_{a{x_1 x_2}(\pm x_3)} \in
I(a_{\tilde{x}_1 \tilde{x}_2}(-\tilde{x}_3), a_{\tilde{x}_1 \tilde{x}_2}(\tilde{x}_3))$ if 
$a_{\tilde{x}_1 \tilde{x}_2}(\tilde{x}_3) > \rho_{\tilde{x}_1 0} > \rho_{\tilde{x}_1 +_r 1, 0} > 
a_{\tilde{x}_1 \tilde{x}_2}(-\tilde{x}_3)$.  As a result, the set of all $i_{a_{x_1 x_2} (\pm x_3)}$'s is allowable 
in $I$.  A similar argument holds for all $j_{b_{x_1 x_2}(\pm y_3)}$'s in $J$.

The subloops $L$ and $M$ are found by using Algorithm~\ref{a1}, and the subloops $QR$ can again be determined by 
Algorithm~\ref{a4} and 5.

Let us consider the converse, i.e., whether any expression of the form shown in Eq.(\ref{3.34}) and satisfying the
five accompanying statements is a term in Eq.(\ref{3.32}).  The only thing we need to do to substantiate this 
converse statement is to  prove that if the set of all $i_{a_{x_1 x_2}(\pm x_3)}$'s is allowable in $I$, the set of
all $j_{b_{x_1 x_2}(\pm y_3)}$'s is allowable in $J$, the set of all
$i_{\rho_{x_1 x_2}}$'s satisfies Eq.(\ref{3.33.1}) and the set of all $j_{\sigma_{x_1 x_2}}$'s satisfies 
Eq.(\ref{3.33.2}), then for each fixed $x_1$, the set of all $i_{a_{x_1 x_2}(\pm x_3)}$'s and 
$j_{b_{x_1 x_2}(\pm y_3)}$'s together with $i_{\rho_{x_1 x_2}}$ and $j_{\sigma_{x_1 x_2}}$ is allowable in 
$f^{I(\rho_{x_1 0}, \rho_{x_1 +_r 1, 0}) J(\sigma_{x_1 +_r 1, 0}, \sigma_{x_1 0})}$.

Indeed, let all the five statements accompanying Eq.(\ref{3.34}) be satisfied.  Obviously the pair
$i_{\rho_{x_1 1}}$ and $j_{\sigma_{x_1 1}}$ forms an allowable set of contraction pairs in the loop 
$I(\rho_{x_1 0}, \rho_{x_1 +_r 1, 0})J(\sigma_{x_1 +_r 1, 0}, \sigma_{x_1 0})$.
Assume that $(i_{\rho_{x_1 1}}$, $j_{\sigma_{x_1 1}}$, $i_{\rho_{x_1 2}}$, 
$j_{\sigma_{x_1 2}}$, \ldots, $i_{\rho_{x_1 x_2}}$, $j_{\sigma_{x_1 x_2}})$ is 
allowable.  If $x_2 < s_{x_1}$, consider the set $(i_{\rho_{x_1 1}}$, 
$j_{\sigma_{x_1 1}}$, $i_{\rho_{x_1 2}}$, $j_{\sigma_{x_1 2}}$, \ldots, 
$i_{\rho_{x_1 x_2 + 1}}$, $j_{\sigma_{x_1 x_2 + 1}})$.  It is clear that
$i_{\rho_{x_1 x_2 +1}} \in I(\rho_{x_1 x_2}, \rho_{x_1 +_r 1, 0})$ 
and $j_{\sigma_{x_1 x_2 + 1}} \in J(\sigma_{x_1 +_r 1, 0}, \sigma_{x_1 x_2})$
from Eqs.(\ref{3.33.1}) and (\ref{3.33.2}).  Hence both $i_{\rho_{x_1 x_2 + 1}}$
and $j_{\sigma_{x_1 x_2 + 1}} \in I(\rho_{x_1 0}, \rho_{x_1 +_r 1, 0})
J(\sigma_{x_1 +_r 1, 0}, \sigma_{x_1 0}) (\rho_{x_1 \tilde{x}_2}, 
\sigma_{x_1 \tilde{x}_2}) \; \forall \; \tilde{x}_2 = 1$, 2, \ldots, and $x_2$.  As a result, 
$(i_{\rho_{x_1 1}}, j_{\sigma_{x_1 1}},$ $i_{\rho_{x_1 2}}, j_{\sigma_{x_1 2}},
\ldots, i_{\rho_{x_1 s_{x_1}}}, j_{\sigma_{x_1 s_{x_1}}})$ is allowable in
$I(\rho_{x_1 0}, \rho_{x_1 +_r 1, 0}) 
J(\sigma_{x_1 +_r 1, 0}, \sigma_{x_1 0})$.

Let us turn to $i_{a_{x_1 x_2}(\pm x_3)}$'s for the fixed $x_1$ we are considering.  Since 
$i_{a_{x_1 x_2}(\pm x_3)} \in I(\rho_{x_1 x_2}, \rho_{x_1, x_2 + 1})$ for $x_2 < s_{x_1}$ or $I(\rho_{x_1 x_2}, 
\rho_{x_1 +_r 1, 0})$ for $x_2 = s_{x_1}$, we have $i_{a_{x_1 x_2}(\pm x_3)} \in 
I(\rho_{x_1 0}, \rho_{x_1 +_r 1, 0}) J(\sigma_{x_1 +_r 1, 0}, \sigma_{x_1 0})(\rho_{x_1 \tilde{x}_2}, 
\sigma_{x_1 \tilde{x}_2})$ for $\tilde{x}_2 \leq x_2$, or $i_{a_{x_1 x_2}(\pm x_3)} \in 
I(\rho_{x_1 0}, \rho_{x_1 +_r 1, 0}) J(\sigma_{x_1 +_r 1, 0}, \sigma_{x_1 0}) 
(\sigma_{x_1 \tilde{x}_2}, \rho_{x_1 \tilde{x}_2})$ for $\tilde{x}_2 > x_2$.  Moreover, the set of all 
$i_{a_{x_1 x_2}(\pm x_3)}$'s for the fixed $x_1$ is allowable in $I$, so this set of $i_{a_{x_1 x_2} (\pm x_3)}$'s 
alone is also allowable in $I(\rho_{x_1 0}, \rho_{x_1 +_r 1, 0}) J(\sigma_{x_1 +_r 1, 0}, \sigma_{x_1 0})$.  
Furthermore, $i_{a_{x_1 x_2}(\pm x_3)} \in I(\rho_{x_1 0}, \rho_{x_1 +_r 1, 0}) 
J(\sigma_{x_1 +_r 1, 0}, \sigma_{x_1 0})(b_{x_1 \tilde{x}_2 (-y_3)}, b_{x_1 \tilde{x}_2 (y_3)})$ for 
$0 \leq \tilde{x}_2 \leq s_{x_r}$ and $(\sigma_{x_1 +_r 1, 0} < b_{x_1 \tilde{x}_2 (-y_3)} < 
b_{x_1 \tilde{x}_2 (y_3)} < \sigma_{x_1 0})$.  A similar argument applies to the set of all 
$j_{b_{x_1 x_2}(\pm y_3)}$'s for the fixed $x_1$.  Consequently, for each fixed $x_1$, the set of all 
$i_{a_{x_1 x_2}(\pm x_3)}, j_{b_{x_1 x_2}(\pm y_3)}, i_{\rho_{x_1 x_2}}, j_{\sigma_{x_1 x_2}}$'s is allowable in 
$I(\rho_{x_1 0}, \rho_{x_1 +_r 1, 0}) J(\sigma_{x_1 +_r 1, 0}, \sigma_{x_1 0})$.  Q.E.D.

\begin{theorem}
There exists an invertible Poisson morphism between the Poisson algebra of Weyl-ordered loop variables and the 
Poisson algebra of normal-ordered loop variables.
\label{t1}
\end{theorem}
{\bf Proof}.  This is a direct consequence of Lemmas~\ref{l9}, \ref{l12} and \ref{l13}. Q.E.D.

\vskip 1pc
\noindent \Large{\bf \hskip .2pc Acknowledgments}
\vskip 1pc
\noindent 
\normalsize
We would like to thank O.T.Turgut for useful discussions.  Indeed, C.W.H.L. is indebted to him for explaining the
content of Ref.\cite{ratu96} in detail.  C.W.H.L. would also like to thank C.Macesanu for his technical help in the 
use of \LaTeX $\;$ at an early stage of this work.  This work was supported in part by the U.S. Department of Energy
under grant DE-FG02-91ER40685.

\end{document}